\documentclass[%
 reprint,
superscriptaddress,
 amsmath,amssymb,
 aps,
 pra,
]{revtex4-1}

\usepackage{graphicx}
\usepackage{dcolumn}
\usepackage{bm}
\usepackage{natbib}
\usepackage{hyperref}
\usepackage{siunitx}
\usepackage[caption=false]{subfig}
\usepackage{booktabs}
\usepackage[export]{adjustbox}
\usepackage{multirow}
\usepackage{physics}
\usepackage{xcolor}
\usepackage{accents}

\newcommand{\zainvec}[1]{\ensuremath{\mathbf{#1}}}
\newcommand{\mm}[1]{\ensuremath{\mathbf{#1}}}
\newcommand{\expens}[1]{\mathbb{E}[\langle {#1} \rangle]}
\newcommand{\expq}[1]{\langle {#1} \rangle}
\newcommand{\exps}[1]{\mathbb{E}[ {#1} ]}

\begin{document}

\title{Multi-mode cooling of a Bose-Einstein condensate with linear quantum feedback }

\def\ANU{Department of Quantum Science and Technology, The Australian National University, Canberra, ACT 2601, Australia}

\author{Zain Mehdi}
 \email{zain.mehdi@anu.edu.au}%
 \affiliation{\ANU}%
\author{Matthew L. Goh}
 \affiliation{Department of Materials, University of Oxford, Parks Road, Oxford OX1 3PH, United Kingdom}%
 \author{Matthew J. Blacker}
 \affiliation{Department of Applied Mathematics and Theoretical Physics, University of Cambridge, Cambridge CB3 0WA, United Kingdom}%
\author{Joseph J. Hope}
 \affiliation{\ANU}%
\author{Stuart S. Szigeti}
 \affiliation{\ANU}%

\date{\today}
\begin{abstract}
We theoretically investigate measurement-based feedback control over the motional degrees of freedom of an oblate quasi-2D atomic Bose-Einstein condensate (BEC) subject to continuous density monitoring. We develop a linear-quadratic-Gaussian (LQG) model that describes the multi-mode dynamics of the condensate's collective excitations under continuous measurement and control. Crucially, the multi-mode cold-damping feedback control we consider uses a realistic state-estimation scheme that does not rely upon a particular model of the atomic dynamics. We present analytical results showing that collective excitations can be cooled to below single-phonon average occupation (ground-state cooling) across a broad parameter regime, and identify the conditions under which the lowest steady-state phonon occupation is asymptotically achieved. Further, we develop multi-objective optimization methods that explore the trade-off between cooling speed and the final energy of the cloud, and provide numerical simulations demonstrating the ground-state cooling of the lowest ten motional modes above the condensate ground state. Our investigation provides concrete guidance on the feedback control design and parameters needed to experimentally realize a feedback-cooled BEC.
\end{abstract}


\maketitle


\section{Introduction}
The cooling of atomic systems into or near their motional ground state underpins quantum science experiments, enabling critical advances in quantum information processing~\cite{Wineland1998a,Schmidt-Kaler2000,Feng2020,Lee2023c,Holzl2023,Fabrikant2024}, optical clocks~\cite{Chou2010,Ludlow2015a,Chen2017}, quantum-enhanced sensing~\cite{Szigeti2020a,Szigeti2021a,Cassens2025}, analogue quantum simulators~\cite{Gaebler2010,Liao2010,Taie2022,Arguello-Luengo2024,Bharti2024}, and tests of fundamental science~\cite{Becker2018,Tino2019,Cacciapuoti2020,Bassi2022}. The needed motional control is well-established in single atom~\cite{Kaufman2012} and single-mode optomechanical systems, including those engineered by placing a cold-atom ensemble in a high-finesse cavity~\cite{Mahajan2013,Kroeger2020a}). However, multi-mode motional control of quantum many-body systems such as degenerate quantum gases remains an open theoretical and experimental challenge.

One pathway to motional control of many-body atomic systems is offered by closed-loop feedback control, wherein motional excitations are damped by optical forces chosen based upon information gained through real-time monitoring of the system. This approach has been exceptionally successful in the ground-state cooling of trapped ions~\cite{Steixner2005,Bushev2006a}, nanoparticles~\cite{Tebbenjohanns2019,Tebbenjohanns2021a,Magrini2021}, and mechanical resonators~\cite{Cohadon1999a,Martin2004,Kleckner2006,Chan2011,Schafermeier2016,Kralj2022}.

The application of feedback cooling to quantum gas experiments holds great potential, offering avenues for generating ultra-large Bose-Einstein condensates (BECs) beyond the capabilities of evaporative cooling~\cite{Mehdi2024a}, preparing highly non-equilibrium many-body steady-states~\cite{Yamaguchi2023,Young2021b}, and generating  entanglement between motional modes~\cite{Wade2015}. Experimentally, feedback control of the motional dynamics of quantum gases has
yet to be realized, however the basic elements required to implement real-time feedback control in cold-atom systems have been demonstrated -- namely, spatially-resolved non-destructive imaging~\cite{Andrews1996,Altuntas2022,Altuntas2023} and high-bandwidth spatiotemporal optical potentials~\cite{Gauthier2016,Gauthier2021,Gajdacz2013,Henderson2009c}. In order to leverage these technological capabilities towards real-time motional control of quantum gases, comprehensive theoretical models are needed to support experimental efforts and motivate control schemes. This is the primary motivation and focus of this work.

Although there have been a number of theoretical studies into the feedback control of ultracold atomic gases~\cite{Wiseman2001,Thomsen2002,Haine2004,Johnsson2005,Szigeti2009,Szigeti2010,Hush2013,Wade2015,Wade2016a,Schemmer2018,Hurst2020,Goh2022a,Yamaguchi2023,Zhu2025}, a complete theoretical understanding of measurement-based feedback cooling in these systems remains lacking. This is partially due to the inherent difficulty in faithfully modelling continuously-monitored quantum gases, which are many-body systems with complex multi-mode structure and non-trivial quantum correlations. Although substantial progress has been made in numerically simulating the full-field dynamics for modest atomic ensembles~\cite{Hush2013,Zhu2025}, these computationally-intensive techniques cannot easily be used for exploring the large parameter spaces of control and measurement variables, and can thus only provide limited insight into the intricate interplay of native atomic dynamics, quantum measurement, state estimation, and spatially-resolved feedback. There is thus a critical need for a modelling approach that delivers immediate yet quantitatively-accurate insights into the viability of a feedback-cooled BEC across a broad parameter regime.

In this work, we develop an analytically-tractable linear quadratic Gaussian (LQG) theory of the low-energy motional dynamics of a BEC under both (1) continuous density monitoring and (2) control via a spatiotemporal potential. Our theoretical framework extends previous LQG models of controlled BECs~\cite{Wade2015,Wade2016a,Schemmer2018,Bouchoule2018} that employed highly-idealized state estimation models, which are not reflective of how feedback control would be implemented in a real experimental system, making them incapable of drawing firm conclusions on the realistic limits of feedback cooling and the stability of the control loop. We build upon these models by explicitly accounting for real-time filtering of the measurement signal in an experimentally-realistic control scheme, which enables us to assess the viability of controlling the motion of a multi-mode BEC system. 

Specifically, we apply our quantum LQG framework to investigate the feasibility of feedback cooling the low-energy motional modes of a cylindrically-symmetric `pancake' BEC to their motional ground state. Our analysis, largely analytical and semi-analytical, reveals a trade-off between cooling speed and the final energy of the feedback-cooled system due to the competition between measurement backaction, signal-to-noise of the measurement signal, and the temporal bandwidth of the control loop. We develop a semi-analytical optimization procedure which allows this trade-off to be balanced by choice of measurement and control parameters, revealing a broad parameter range in which collective excitations can be controlled to their motional ground state in tens of trap periods. Our analysis demonstrates that multi-mode ground-state cooling in BEC systems is achievable with parameters accessible in modern ultracold-atomic experimental apparatus, strengthening the conclusions of prior investigations~\cite{Szigeti2010,Hush2013,Mehdi2024a,Zhu2025} and paving the way for the near-term experimental realization of a feedback-cooled BEC.


\section{Background theory: Dispersively-monitored BECs}
\begin{figure*}
	    \centering
    \includegraphics[width=\textwidth]{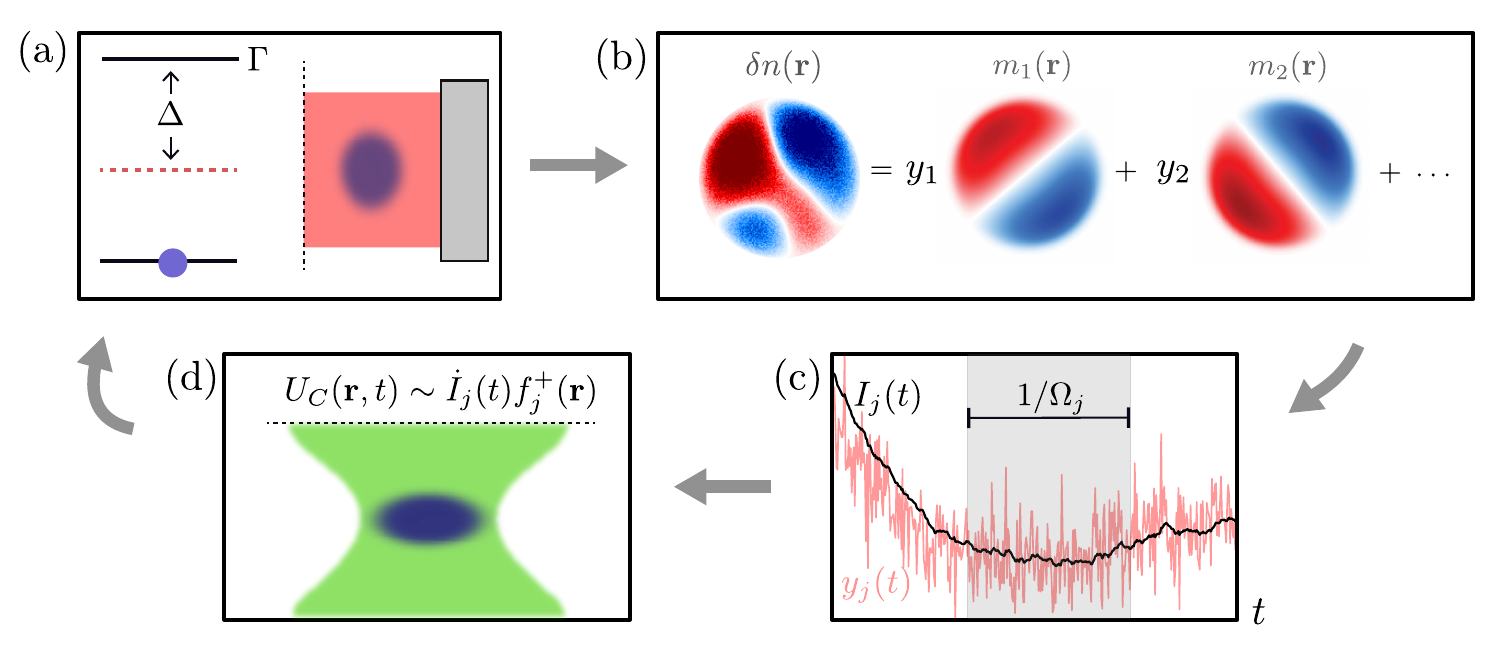}
    \caption{Schematic of the feedback control loop. (a) Non-destructive imaging of the atomic density is performed by continuous illumination of the atomic cloud with a coherent light field far detuned from resonance, i.e. $\Delta\gg \Gamma$. (b) Measured fluctuations of the atomic density are decomposed onto a finite basis of mode functions, $m_j(\mathbf{r})$ [see Eq.~\eqref{eq:linearizedMeasurementOperator}]. (c) The projected measurement current associated with the $j$th mode, $y_j(t)$, is constructed from the time series of many measurements (red). High-frequency noise is removed from the measurement current associated with each mode, using a low-pass filtered with bandwidth $\Omega_j$ (grey shading indicates a temporal averaging window of $1/\Omega_j$). (d) The derivative of low-pass-filtered measurement current, $I_j(t)$, is used to inform the control, which is actuated by a high-bandwidth spatiotemporal optical-dipole potential (green). Specifically, we consider a multi-mode cold-damping protocol realised by choosing the spatiotemporal control potential to be linear combinations of the Bogoliubov $f^+_j(\mathbf{x})$ functions (see Eq.~\eqref{eq:SpatiotemporalPotential_ModeBasis}) with gain proportional to the time derivative of the filtered measurement current, $\dot{I}_j(t)$.  }
    \label{fig:Schematic}
\end{figure*}

We consider a realistic feedback control scheme based on dispersive optical imaging of an oblate (quasi-2D) BEC~\cite{Szigeti2009, Szigeti2010,Mehdi2024a}, represented in Fig.~\ref{fig:Schematic}. In this scheme, weak measurements of the atomic density are performed by stroboscopically illuminating the atomic cloud along its tightly-trapped axis (taken to be the $z$ direction) by pulses of coherent light far-detuned from atomic resonance. An estimate of the atomic density can be extracted from the phase of the scattered light, which is used to inform a real-time controller aimed at damping observable density of the atomic cloud. \par 

Below we briefly review the theory of BECs subject to continuous dispersive monitoring. We begin with a brief review of the semiclassical description of dispersive imaging (where the optical field is treated classically), which illustrates the measurement process conceptually. We then review the full-field quantum measurement theory, which includes key effects arising from the quantization of the probe field: measurement backaction (heating) of the BEC system, and quantum projection noise in the measurement output. Both of these effects are crucial in order to make quantitative predictions of feedback control in low-energy quantum gases~\cite{Szigeti2009,Szigeti2010,Mehdi2024a}.

\subsection{Dispersive imaging: semiclassical description}
Treating the light field classically, we can describe this measurement protocol in terms of the complex polarisability of an ensemble of two-level atoms with excited-state linewidth $\Gamma$, which gives the following refractive index for the atomic cloud~\cite{Knight1983c,Ketterle1999_Arxiv}:
\begin{align}
\label{eq:RefractiveIndexAtomicCloud}    n_{\rm ref}(\mathbf{r}) \approx 1 + \rho(\mathbf{r})\frac{\sigma_0 \lambda }{4\pi}\left(\frac{i}{1+\delta^2}+\frac{\delta}{1+\delta^2}\right) \,,
\end{align}
where $\rho(\mathbf{r})$ is the atomic density, $\sigma_0 = 3\lambda^2/(2\pi)$ is the resonant absorption cross-section of the light field given an atomic transition frequency $\omega_0=2\pi c/\lambda$, and $\delta = \Delta/(\Gamma/2)$ is the dimensionless detuning of the light field with respect to the atomic transition -- i.e. $\Delta = \omega_0-\omega_{\rm laser}$. The real and imaginary parts of Eq.~\eqref{eq:RefractiveIndexAtomicCloud} respectively describe dispersion and absorption of scattered light by the atomic medium. For far-detuned light, i.e. $\delta \gg 1$, the dispersive term dominates in Eq.~\eqref{eq:RefractiveIndexAtomicCloud} such that the scattered light field accumulates a spatially-dependent phase in the $xy$ plane: $\phi(\mathbf{x})\approx\bar{n}(\mathbf{x})\sigma_0/(2\delta)$, where $\bar{n}(\mathbf{x})=\int dz \rho(\mathbf{x},z)$ is the two-dimensional column density in terms of the 2D coordinates $\mathbf{x}=(x,y)$. The 2D atomic density $\bar{n}(\mathbf{x})$ can then be estimated by a spatially-resolved measurement of the optical phase, which can be achieved by interfering the scattered and unscattered components of the light field. This forms the basis of well-established dispersive imaging techniques which have been used to take multiple \emph{in situ} images of a single condensate and track its motion in real time with minimal disturbance due to spontaneous emission~\cite{Andrews1996,Bradley1997,Gajdacz2013,Everitt:2017,Wigley2016a}. In this manuscript we will focus on the phase-contrast imaging technique, which can be modelled as homodyne detection of the scattered light, where the unscattered light is treated as a local oscillator for the homodyne detection.

Although the above semiclassical description of dispersive imaging is insightful, it neglects quantum correlations of the optical field as well as the multi-mode nature of the atomic cloud, which are essential in order to describe the backreaction on the atomic cloud from each measurement. In order to correctly model these effects, and their impact on the spatiotemporal control of the BEC system, we must use a full-field quantum theory of dispersive measurements. 

\subsection{Quantum theory of a continuously-monitored BEC}
The full-field theory of non-destructive dispersive measurements in a BEC system has been well-developed using the framework of quantum continuous measurements~\cite{Dalvit2002b,Szigeti2009,Szigeti2010}. Here the optical field is treated quantum mechanically as a Markovian reservoir, after formally eliminating the atomic and optical degrees of freedom along the imaging axis ($z$) and making typical rotating-wave and Born-Markov approximations~\cite{Szigeti2009,Szigeti2010,MehdiPhD2024}. The result is a stochastic master equation (SME) describing the conditional evolution of the many-body atomic state associated with a particular measurement record.

We consider the case of quasi-continuous monitoring, wherein the rate of stroboscopic probing is taken to be much faster than the atomic dynamics, such that the record of measurement results can be described in terms of the \emph{measurement current} (in It\^o form)~\cite{Szigeti2009,MehdiPhD2024}:
\begin{align}
\label{eq:MeasurementCurrent}
	dy(\mathbf{x},t) = 2\sqrt{\alpha\eta}\langle \hat{M}(\mathbf{x})\rangle dt + dW(\mathbf{x},t)
\end{align}
where $\mathbf{x}=\{x,y\}$ are coordinates in the imaging plane, $\alpha$ is a positive real number that parameterizes the strength of the measurement (units of area/time), and $\eta\in (0,1]$ is the measurement efficiency such that $\eta=1$ corresponds to the case of perfect detection considered thus far, and $\eta=0$ corresponds to no detection of the scattered light. 

The measurement current, Eq.~\eqref{eq:MeasurementCurrent}, has two contributions. First, a deterministic term that is proportional to the phase of the scattered light, which gives a `blurred' estimate of the atomic density $\hat{n}(\mathbf{x})=\hat{\psi}^\dag(\mathbf{x})\hat{\psi}(\mathbf{x})$, i.e. 
\begin{align}
\label{eq:MeasOperatorDef}
\hat{M}(\mathbf{x}) &= \int d^2\mathbf{x'}\; \hat{n}(\mathbf{x}) K(\mathbf{x-x'}) \,,
\end{align}
where the convolution with the kernel function
\begin{align}
\label{eq:MeasKernelDef}
    K(\mathbf{x}) = \frac{1}{(2\pi)^2} \int  d^2\mathbf{k}\;e^{i\mathbf{k}\cdot\mathbf{x}}e^{-|r_D \mathbf{k}|^4/16}
\end{align}
encodes blurring of the atomic density that arises due to the minimum divergence of the light field over the cloud's spatial extent along the imaging axis. This sets a minimum achievable imaging resolution scale $r_{\rm D} = \sqrt{R_z/k_0}$~\cite{Szigeti2009,Dalvit2002b,MehdiPhD2024}, where $k_0$ is the wavevector of the imaging laser and $R_z$ is the $2\sigma$ width of the gas along the imaging axis. In this work, we will study the control of low-energy excitations above the BEC ground state, which have characteristic lengthscales ($\sim1-10\mu$m) much larger than $r_D$ ($\lesssim 1\mu$m). As a result, the blurring of the measured atomic density can be neglected in the effective low-energy theory we will derive in the following section (see Eqs.~\eqref{eq:M_approx_constant}-\eqref{eq:linearizedMeasurementOperator}). 

The second term in Eq.~\eqref{eq:MeasurementCurrent} describes the quantum projection noise associated with continuous measurement of the scattered light's phase. The Wiener process $ dW(\mathbf{x},t)$ is a field of real-valued Gaussian noises with zero mean and correlations $\expens{dW(\mathbf{x},t)dW(\mathbf{y},t)}=\delta^{(2)}(\mathbf{x}-\mathbf{y})dt$. Should the measurement strength $\alpha$ or detection efficiency $\eta$ be made vanishingly small, the measurement current itself approaches a Wiener process -- that is, if $\alpha=0$ or $\eta=0$, the measurement record contains no information of the atomic state, as expected.

Given a particular measurement record, $dy(\mathbf{x},t)$, the conditional dynamics of the many-body atomic system is described by the (It\^o) SME~\cite{Szigeti2009,Szigeti2010,MehdiPhD2024}:
\begin{align}
\label{eq:SME_2D}
	d\hat{\rho}_c= &-\frac{i}{\hbar}[\hat{H},\hat{\rho}_c]dt+\alpha \int d^2\mathbf{x}\:\mathcal{D}[\hat{M}(\mathbf{x})]\hat{\rho}_c dt  \\ \notag &+ \sqrt{\alpha\eta }\int d^2\mathbf{x}\:\mathcal{H}[\hat{M}(\mathbf{x})]\hat{\rho}_c dW(\mathbf{x},t)  \,.
\end{align}
In addition to the `native' Hamiltonian evolution given by $\hat{H}$, Eq.~\eqref{eq:SME_2D} describes measurement-induced motional dephasing (i.e. decoherence) at a rate proportional to $\alpha$ in terms of the Lindbladian superoperator:
\begin{align}
    \mathcal{D}[\hat{L}]\hat{\rho}_c =  \hat{L}\hat{\rho}_c\hat{L}^\dag - \frac{1}{2}\left(\hat{L}^\dag\hat{L}\hat{\rho}_c+\hat{\rho}_c\hat{L}^\dag\hat{L}\right),
\end{align}
as well as `innovations' due to the knowledge obtained from the measurements record, with:
\begin{align}
    \mathcal{H}[\hat{L}]\hat{\rho}_c = \hat{L}\hat{\rho}_c+\hat{\rho}_c\hat{L}-\textrm{Tr}\left[(\hat{L}+\hat{L}^\dag)\hat{\rho}_c\right]\,.
\end{align}
The innovations term in Eq.~\eqref{eq:SME_2D} is additionally a function of the measurement efficiency $\eta\in [0,1]$ -- if $\eta=0$, the innovations term vanishes as there is no `information' with which to update the conditional quantum state. Note that $dW(\mathbf{x},t)$ in this equation is precisely the same quantum projection noise as in Eq.~\eqref{eq:MeasurementCurrent}. 

\subsection{Conditional and unconditional expectation values}
Equation~\eqref{eq:SME_2D} describes the evolution of the \emph{conditional} atomic state $\hat{\rho}_c$ -- that is, the atomic state associated with a particular measurement record (i.e. a single experimental run). Correspondingly, we will use the notation $\expq{\star}$ to refer to expectation values with respect to the conditional state, i.e.
\begin{align}
    \expq{\hat{A}} \equiv \textrm{Tr}\left[\hat{A}\hat{\rho}_c\right]
\end{align}
for some operator $\hat{A}$. In order to obtain the \emph{unconditioned} expectation value, we must take an ensemble average over a large number of experimental realisations. We denote this ensemble averaging using $\exps{\star}$, such that the unconditional expectation value of some operator $\hat{A}$ is given by 
\begin{align}
    \expens{\hat{A}} \equiv \frac{1}{N_{\rm S}}\sum_{i=1}^{N_{\rm S}} \expq{\hat{A}}^{(i)} \,,
\end{align}
where $\expq{\hat{A}}^{(i)}$ denotes the conditional expectation value of the $i$th experimental run, and $N_{\rm S}\gg 1$ is the number of samples in the ensemble.

\subsection{Hamiltonian evolution of a quasi-2D BEC}
In the case where the system Hamiltonian $\hat{H}$ is independent of the measurement record, ensemble averaging over conditional trajectories gives a Lindbladian master equation for the unconditional state $\hat{\rho}\equiv \exps{\hat{\rho}_c}$, i.e.
\begin{align}
\label{eq:PureDephasing_NoFeedback}
	\frac{d}{dt}\hat{\rho}= &-\frac{i}{\hbar}[\hat{H},\hat{\rho}]+\alpha \int d^2\mathbf{x}\:\mathcal{D}[\hat{M}(\mathbf{x})]\hat{\rho}    \,,
\end{align}
where we have used the property that $\hat{\rho}_c$ is a non-anticipating function of $t$ to compute $\exps{\hat{\rho}_c dW(\mathbf{x},t)}= \hat{\rho}_c \exps{dW(\mathbf{x},t)} = 0$. However, the equivalence between the ensemble-averaged SME and Lindbladian evolution is broken in the case of closed-loop feedback, where the system is controlled by a term in the Hamiltonian that depends explicitly on the conditional state. In this case, $\exps{[\hat{H},\hat{\rho}_c]}\neq [\hat{H},\hat{\rho}]$, such that Eq.~\eqref{eq:PureDephasing_NoFeedback} no
longer describes the average dynamics of the atomic system -- instead, there will
be additional terms describing the mean value of the control.

In this work, we consider Hamiltonian evolution described by $\hat{H}=\hat{H}_{\rm atom}+\hat{H}_{\rm fb}(t)$, which contains contributions from the dimensionally-reduced atomic dynamics in the $xy$ plane
\begin{align}
\label{eq:AtomicHamiltonian}
    \hat{H}_{\rm atom} &= \int d^2\mathbf{x}\left(  \hat{\psi}^\dag(\mathbf{x})h_0(\mathbf{x})\hat{\psi}(\mathbf{x}) + \frac{g_{\rm 2D}}{2} \hat{\psi}^\dag(\mathbf{x})^2\hat{\psi}(\mathbf{x})^2\right) \,,
\end{align}
as well as a general control term $\hat{H}_{\rm fb}(t)$, which is a function of the conditional quantum state of the atoms. Here, $h_0(\mathbf{x})=-\hbar^2\nabla_{\rm 2D}^2/(2m)+m\omega_0^2(x^2+y^2)/2$ is the single-particle Hamiltonian for a cylindrically-symmetric 2D harmonic oscillator, and $g_\text{2D}$ is the effective 2D interaction strength related to the inter-atomic scattering length $a_s$ and the gas width $R_z$:
\begin{equation}
g_\text{2D}=\frac{2\sqrt{2\pi}\hbar^2 a_s}{m R_z} \,.
\end{equation}
This result is consistent with assuming the atomic density profile along the tightly-trapped axis is well approximated by a Gaussian of the form $e^{-(z/R_z)^2}$~\cite{Pethick2008}.

In reporting numerical quantities, we work in a natural system of units set by the harmonic oscillator frequency -- specifically, energies are expressed in units of $E_0=\hbar\omega_0$, and length scales in units of $l_0 = \sqrt{\hbar/(m\omega_0)}$.

\section{LQG Theory of a feedback-controlled BEC\label{sec:LGTheory}}
In this section, we develop an analytically and numerically-tractable model of continuous measurement and feedback of a Bose gas in the low-energy regime. This model is constructed using a perturbative approach, where Eq.~\eqref{eq:SME_2D} is linearized in terms of fluctuations around the BEC ground state -- in the absence of measurement, this reveals the collective excitation spectrum of a quasi-2D BEC. We find that the linearization of measurement process describes continuous monitoring of each collective excitation, with correlated noises between different phonon modes. Then, under a Gaussian approximation, the dynamics of the Bose gas under continuous measurement and feedback can be written in terms of a discrete set of equations for first- and second-order moments of the collective excitations.

\subsection{The conditional Bogoliubov approach}
In order to linearise the SME \eqref{eq:SME_2D}, we adopt the symmetry-breaking Bogoliubov approach, where the field operator is decomposed as
\begin{align}
\label{eq:Bog_Decomp}
    \hat{\psi}(\mathbf{x}) = \phi_0(\mathbf{x}) + \hat{\delta}(\mathbf{x})
\end{align}
where $\phi_0(\mathbf{x})$ is identified as the condensate ground-state wavefunction (normalized to the number of condensate atoms $N_c$) and $\hat{\delta}(\mathbf{x})$ describes fluctuations about this ground-state. In our analysis we will restrict ourselves to the low-energy regime, where $\hat{\delta}(\mathbf{x})$ is treated as small with respect to $\phi_0(\mathbf{x})$. 

Typically in Bogoliubov analysis, the condensate wavefunction is defined by the mean of the field operator, $\phi_0(\mathbf{x}) = \langle \hat{\psi}(\mathbf{x}) \rangle$, immediately implying $\langle \hat{\delta}(\mathbf{x}) \rangle = 0$. However, for \emph{conditional} quantum state evolution, we instead have 
\begin{align}
\label{eq:condensateMF_def}
    \phi_0(\mathbf{x}) \equiv \mathbb{E}[\langle \hat{\psi}(\mathbf{x}) \rangle ] \,,
\end{align}
where $\exps{\star}$ denotes an average over conditional trajectories, i.e. over many experimental runs. In this case, the conditional expectation of the fluctuation operator will in general be non-zero $\langle \hat{\delta}(\mathbf{x}) \rangle\neq 0$. Note that Eq.~\eqref{eq:condensateMF_def} neglects the contribution of the thermal cloud to $\mathbb{E}[\langle \hat{\psi}(\mathbf{x}) \rangle ]$, which is appropriate for the low-energy regime considered in this work, which is characterized by weak fluctuations around the condensate mode.

Substituting Eq.~\eqref{eq:Bog_Decomp} into Eq.~\eqref{eq:AtomicHamiltonian} and retaining terms up to quadratic order in $\hat{\delta}$ gives an approximate atomic Hamiltonian that is diagonalized by the Bogoliubov transformation $\hat{\delta}(\mathbf{x}) = \sum_j \left( u_j(\mathbf{x}) \hat{b}_j - v_j^*(\mathbf{x}) \hat{b}_j^\dag \right)$~\cite{Pethick2008}:
\begin{equation}
\label{eq:Ham_Bog_quasiparticles}
	\hat{H}_{\rm atom} \approx \sum_{j}  \hbar\omega_j \left(\hat{b}_j^\dag \hat{b}_j+\frac{1}{2}\right) \,.
\end{equation}
The operators $\hat{b}_j$ and $\hat{b}_j^\dag$ describe the annihilation and creation of \emph{quasi-particle} (collective) excitations, satisfying the bosonic commutation relation $[\hat{b}_j, \hat{b}^\dag_k]=\delta_{jk}$. The corresponding energies $\hbar \omega_j$ and eigenmodes $u_j(\mathbf{x}),v_j(\mathbf{x})$ are determined by solving the Bogoliubov-de-Gennes equations~\cite{Pethick2008}: 
\begin{align}
\label{eq:BdGequations}
	\begin{pmatrix}
		\mathcal{L}_{\rm GP}+g_{\rm 2D} n_0(\mathbf{x}) && -g_{\rm 2D} \phi_0(\mathbf{x})^2 \\
		-g_{\rm 2D} \phi_0^*(\mathbf{x})^2 && \mathcal{L}_{\rm GP} +g_{\rm 2D} n_0(\mathbf{x})
	\end{pmatrix}&
	\begin{pmatrix}
		u_j(\mathbf{x}) \\ v_j(\mathbf{x})
	\end{pmatrix} \\ \notag &=
	\omega_j 
	\begin{pmatrix}
		u_j(\mathbf{x}) \\ -v_j(\mathbf{x})
	\end{pmatrix},
\end{align}
where $\mathcal{L}_{\rm GP} = h_0(\mathbf{x}) +  g_{\rm 2D} n_0(\mathbf{x}) - \mu $. Here $n_0(\mathbf{x})=|\phi_0(\mathbf{x})|^2$ is the condensate density, and $\mu$ is the corresponding zero-temperature chemical potential. In general, these modes are not exactly orthogonal to $\phi_0(\mathbf{x})$, however this can be achieved with the appropriate projector (see Ref.~\cite{Blakie2008}).

For the linearization of the measurement terms in Eq.~\eqref{eq:SME_2D}, it will be more convenient to work with quadratures of the quasi-particle operators: 
\begin{subequations}
\label{eq:AtomicQuadratures}
	\begin{align}
	\hat{X}_j 	&= \frac{1}{\sqrt{2}}\left( \hat{b}_j + \hat{b}_j^\dag \right)\,, \\
	\hat{Y}_j 	&= \frac{i}{\sqrt{2}} \left( \hat{b}_j^\dag - \hat{b}_j \right)\,,
\end{align}
\end{subequations}
which satisfy the (dimensionless) position-momentum commutation relation: $[\hat{X}_j,\hat{Y}_k]=i\delta_{jk}$. The quadratures $\hat{X}_j$ and $\hat{Y}_j$ can be interpreted as the amplitudes of density and phase fluctuations, respectively, for the $j$th collective mode (see Eq.~\eqref{eq:linearizedMeasurementOperator}, below). The mode functions corresponding to these quadratures are given by $f_j^\pm(\mathbf{x}) = \frac{1}{\sqrt{2}}\left( u_j(\mathbf{x}) \pm v_j^*(\mathbf{x}) \right)$, 
in terms of which the linearized atomic Hamiltonian \eqref{eq:Ham_Bog_quasiparticles} becomes:
\begin{equation}
	\hat{H}_{\rm atom} = \hbar\sum_{j=1}\frac{\omega_j}{2}\left( \hat{X}_j^2 + \hat{Y}_j^2 \right) \,.\label{eq:Ham_Bog}
\end{equation}

\subsection{Linearization of the measurement operator}
To derive a linearized form of the conditional SME \eqref{eq:SME_2D}, we first note that Eq.~\eqref{eq:SME_2D} is invariant under the transformation $\hat{M}(\mathbf{x})\rightarrow \hat{M}(\mathbf{x})-f(\mathbf{x})$ in the measurement terms, where $f(\mathbf{x})$ is an arbitrary real-valued function. This allows us to write the SME as
 \begin{align}
\label{eq:SME_DensityFluctuations}
	d\hat{\rho}_c =&-\frac{i}{\hbar}[\hat{H},\hat{\rho}_c]dt+\alpha \int d^2\mathbf{x}\:\mathcal{D}[\delta\hat{M}(\mathbf{x})]\hat{\rho}_c dt \\ \notag &+ \sqrt{\alpha\eta }\int d^2\mathbf{x}\:\mathcal{H}[\delta\hat{M}(\mathbf{x})]\hat{\rho}_c dW(\mathbf{x},t)  \,,
\end{align}
where $\delta\hat{M}(\mathbf{x}) \equiv \hat{M}(\mathbf{x})-n_0(\mathbf{x})$ is the operator describing measured density fluctuations around the condensate density profile $n_0(\mathbf{x})$.
Then, substituting the decomposition of the field operator and retaining only those terms linear in $\hat{\delta}(\mathbf{x})$, we find
\begin{align}
	\delta\hat{M}(\mathbf{x})	&\approx  \left(\phi_0^*(\mathbf{x}) \hat{\delta}(\mathbf{x})+ \phi_0(\mathbf{x}) \hat{\delta}^\dag(\mathbf{x})\right)  * K(\mathbf{x})  \label{eq:M_approx_constant} \\\nonumber
	&\approx \sum_j \left(m_j^-(\mathbf{x}) \hat{X}_j + i m_j^+(\mathbf{x}) \hat{Y}_j\right)* K(\mathbf{x}) \,,
\end{align}
where $m_j^\pm(\mathbf{x}) \equiv \phi^*(\mathbf{x})  f_j^\pm(\mathbf{x}) \mp c.c.$ and we have employed the shorthand for the 2D convolution $F(\mathbf{x})*G(\mathbf{x}) \equiv \int d^2\mathbf{x'}F(\mathbf{x'})G(\mathbf{x-x'})$. For a harmonically-trapped gas, the condensate wavefunction and the mode functions $f^\pm_j(\mathbf{x})$ can be taken to be real (due to the symmetry of the potential, $V(\mathbf{x})=V(-\mathbf{x})$), which implies $m_j^+(\mathbf{x})=0$. Furthermore, the mode functions $m_j(\mathbf{x})$ vary slowly across the spatial extent of the condensate, and thus characteristic lengthscales for low-energy excitations ($\sim1-10\mu$m) can be assumed to be much larger than the imaging resolution scale $r_D$ ($\lesssim 1\mu$m). We may therefore take the limit of $r_D \rightarrow 0$ in Eq.~\eqref{eq:MeasKernelDef}, in which case we have $K(\mathbf{x})\approx \delta(\mathbf{x})$. Therefore, the measurement term may be expressed to linear order as
\begin{equation}
\label{eq:linearizedMeasurementOperator}
	\delta\hat{M}(\mathbf{x})	\approx \sum_j m_j(\mathbf{x}) \hat{X}_j, 
\end{equation}
where $m_j(\mathbf{x}) \equiv m_j^-(\mathbf{x}) = 2 \phi_0(\mathbf{x}) f_j^{-}(\mathbf{x})$ is the mode function describing density fluctuations of the $j$th mode. Substituting this decomposition into Eq.~\eqref{eq:MeasurementCurrent} then provides the linearized measurement current:
\begin{align}
    dY(\mathbf{x},t)  &\approx 2\sqrt{\alpha \eta}\sum_j m_j(\mathbf{x}) \langle \hat{X}_j \rangle + dW(\mathbf{x} ,t)  \,,
\end{align}
which can be represented in the collective mode basis by projecting $dY(\mathbf{x},t)$ onto the basis $\{m_j(\mathbf{x})\}$ and integrating out spatial degrees of freedom, i.e.
\begin{align}
	dy_k(t) &= \int d^2\mathbf{x}\; m_k(\mathbf{x}) dY(\mathbf{x} ,t) \\
	&= 2\sqrt{\alpha \eta}\sum_j \mathcal{M}_{jk} \langle \hat{X}_j \rangle + d\xi_k(t)	  \,.
\end{align}
Here we have defined the vector of Gaussian random noises, $d\xi_k(t)	\equiv \int d^2\mathbf{x}\, m_k(\mathbf{x}) dW(\mathbf{x},t)$,
which have zero mean (as $dW$ has zero mean) and colored correlations $\overline{d\xi_j(t) d\xi_k(t)} = \mathcal{M}_{jk} dt$, where 
\begin{align}
\mathcal{M}_{jk}	\equiv \int d^2\mathbf{x} \; m_j(\mathbf{x}) m_k(\mathbf{x})	\,,
\end{align}
are coefficients describing measurement-induced couplings between modes $j$ and $k$. These couplings can be succinctly written in terms of the matrix $\mm{M}$, 
\begin{align}
    \label{eq:MeasCouplingMatrix}
	[\mm{M}]_{jk}	&= \begin{pmatrix} 
							 \mathcal{M}_{jk} && 0 \\ 
							0 && 0 
						       \end{pmatrix} \,, 
\end{align}
allowing us to express the projected measurement signal as
\begin{align}
    \label{eq:MeasSignal_dy}
    d\mathbf{y}(t) = 2\sqrt{\alpha \eta} \mm{M} \expq{\hat{\zainvec{x}}} dt + \mm{L}d\zainvec{w}(t) \,.
\end{align}
Here the vector $\mathbf{\hat{x}} \equiv \left(\hat{X}_1,\hat{Y}_1,\hat{X}_2,\hat{Y}_2,\dots \right)^\intercal$ describes the means of the mode quadratures, $\mm{L}$ is the Cholesky decomposition of the measurement-coupling matrix $\mm{M}$ (i.e. $\mm{M}=\mm{L}\mm{L^\intercal}$), and 
\begin{align}
\label{eq:MeasNoiseProjected}
	d\zainvec{w}(t)=\left(dw_1(t),0,dw_2(t),0\dots \right)^\intercal
\end{align}
 is a vector of real-valued Gaussian random variables with zero mean and correlations $\exps{dw_j(t)dw_k(t')}=\delta_{jk}\delta(t-t')dt$, representing the quantum projection (backaction) noise driving each mode. Equation~\eqref{eq:MeasSignal_dy} demonstrates that, in the linear regime, continuous monitoring of the atomic density is equivalent to continuous `position' monitoring of the quasi-particle modes.
 

\subsection{Multi-mode quantum Gaussian states\label{sec:GaussianStateAssumption}}
Although the linearization described above significantly simplifies the evolution of the quantum state, we have not yet placed any constraints on the quantum state itself. Therefore, to further simplify our model, we treat the quantum state of the quasi-particle modes as a quantum Gaussian state; one fully characterized by its means and covariances~\cite{Weedbrook2012}. This is an appropriate assumption for our analysis, as we will be considering control of an initial state of thermally-populated quasi-particle modes with the aim of driving the system towards its multi-mode ground state. Both the initial state and the target final state are well-approximated quantum Gaussian in the near-equilibrium regime where the symmetry-breaking Bogoliubov approach, Eq.~\eqref{eq:Bog_Decomp}, is valid. 

Under the assumption of a quantum Gaussian state, the system is fully characterized by the conditional means of the mode quadratures, $\expq{\zainvec{\hat{x}}}$, and the symmetrized covariance matrix:
\begin{align}
\label{eq:SymmetrisedCovarianceMatrix}
    \mm{V} &\equiv \frac{1}{2}\expq{\zainvec{\hat{x}}\zainvec{\hat{x}}^\intercal + (\zainvec{\hat{x}}\zainvec{\hat{x}}^\intercal)^\intercal}-\expq{\zainvec{\hat{x}}}\expq{\zainvec{\hat{x}}^\intercal} \,.
\end{align}
We also introduce the block-diagonal symplectic matrix
\begin{equation}
	\mm{\Sigma}	\equiv \bigoplus_{j=1}^M \begin{pmatrix} 0 && 1 \\ -1 && 0 \end{pmatrix} \,,
\end{equation}
which encodes the canonical commutation relations between the quadratures ($[\hat{X}_j,\hat{Y}_k] = i \delta_{jk}$ and $[\hat{X}_j,\hat{X}_k] = [\hat{Y}_j,\hat{Y}_k] = 0$) via the relation $\hat{\mathbf{x}} \hat{\mathbf{x}}^\intercal - \hat{\mathbf{x}}^\intercal \hat{\mathbf{x}} = i \mm{\Sigma}$. \par

We can then find equations of motion for $\expq{\zainvec{\hat{x}}}$ and $\mm{V}$ directly from Eq.~\eqref{eq:SME_2D}, using the linearized measurement operator \eqref{eq:linearizedMeasurementOperator}. The full details of this calculation are provided in Appendix~\ref{app:LQG_Derivation}; here we will provide only the final result.

Defining the following matrices,
\begin{subequations}
\begin{align}
    \mathbf{A}	&= \bigoplus_{j=1}^M \begin{pmatrix} \omega_j && 0 \\ 0 && \omega_j \end{pmatrix} \,, \\
    	\mm{D}	&= \mm{\Sigma} \mm{M} \mm{\Sigma}^\intercal \,,
\end{align}
\end{subequations}
we may express the evolution of the means and covariances under continuous measurement in vectorized form as:
\begin{subequations}
\label{eq:GaussianEvolution}
\begin{align}
    \label{eq:means_nofb}
	d \langle \hat{\zainvec{x}} \rangle &= \mathbf{\Sigma} \mm{A} \langle \hat{\zainvec{x}} \rangle dt + 2\sqrt{ \alpha \eta} \mm{V} \mm{L} d\zainvec{w} \, \\
	\label{eq:Covariances}
	\dot{\mm{V}} &= (\mm{\Sigma} \mm{A}) \mm{V} + \mm{V} (\mm{\Sigma} \mm{A})^\intercal + \alpha\mm{D} - 4 \alpha \eta \mm{V} \mm{M} \mm{V}\,.
\end{align}
\end{subequations}
This is the standard form for LQG systems subject to continuous monitoring -- see, for example, Chapter~6 of Ref.~\cite{WisemanMilburn2009}. Note that the evolution of the covariance matrix is deterministic, taking the form of a differential Riccati equation that is amenable to analytic treatment. In comparison, the evolution of the means is explicitly stochastic, conditioned on the measurement record by the vector of white noise processes $d\zainvec{w}$ -- note that this is precisely the same noise process as in the measurement signal, Eq.~\eqref{eq:MeasSignal_dy}. 

In the case of a single mode, Eq.~\eqref{eq:GaussianEvolution} describes the dynamics of a continuously-monitored quantum harmonic oscillator, which has been studied in great depth, both for atomic~\cite{Szigeti2013a} and optomechanical systems~\cite{Genes2008}. This close correspondence allows us to adapt well-established control schemes for these single-mode systems to the cooling and motional stabilization of the multi-mode BEC system.

\subsection{Feedback control from spatiotemporal potentials optical \label{sec:ControlPotential}}
Thus far, we have not included the feedback in our description of the linearized atomic dynamics under measurement. In our proposed feedback scheme, the atomic cloud is controlled by a spatiotemporal optical potential actuated by a high-bandwidth spatial-light modulator (SLM), e.g. using a digital micromirror device~\cite{Gauthier2016,Gauthier2021}, painted potentials~\cite{Henderson2009c}, or holographic techniques~\cite{Gaunt2012}. Experimentally, such optical control has been demonstrated with submicron spatial resolution and switching speeds up to $20$kHz~\cite{Gauthier2016,Gauthier2021}. The spatiotemporal structure of low-energy excitations in typical BEC experiments falls well within the bandwidth of the imaging, with relevant lengthscales of tens of microns, and timescales of tens to hundreds of Hz. We are therefore justified in neglecting the bandwidth limits of the control actuation on the control of low-energy excitations, which we will assume for the remainder of this work.

We consider a feedback control actuated by a general spatiotemporal potential, $U_C(\mathbf{x},t)$, described by the Hamiltonian term:
\begin{align}
    \hat{H}_\text{fb}(t) &= \int d^2\mathbf{x}\; U_C(\mathbf{x},t)\hat{\psi}^\dag(\mathbf{x})\hat{\psi}(\mathbf{x}) \,.
\end{align}
In the feedback we will consider in the next section, the control potential will be proportional to the derivative of the measurement signal, which in turn will be approximately linear in the quadrature operators (see Eq.~\eqref{eq:MeasSignal_dy}). Thus, to ensure the Hamiltonian remains quadratic in the quadrature operators and their expectation values, $\hat{H}_\text{fb}(t) $ should only be expanded to linear order in $\hat{\delta}(\mathbf{x})$ upon substitution of Eq.~\eqref{eq:Bog_Decomp}~\footnote{The next order correction Eq.~\eqref{eq:LinearizedFBPotential} is suppressed by a factor of $N_0^{-1/2}$, where $N_0 = \int d^2\mathbf{x}\;n_0(\mathbf{x})$. Thus, the linearization of $\hat{H}_\text{fb}(t)$ is justified for typical condensate with $N_0\gtrsim \mathcal{O}(10^4)$}. This gives a control Hamiltonian of the form (discarding non-operator valued shifts to $\hat{H}$, which may be time-dependent):
\begin{align}
    \hat{H}_\text{fb}(t) &\approx 2\sum_j \hat{X}_j\int d^2\mathbf{x}\; U_C(\mathbf{x},t) \phi_0(\mathbf{x}) f^-_j(\mathbf{x}) \,.
\end{align}

In order to facilitate individual control of each mode, we will take our control potential to be of the following form~\cite{Wade2016a}:
\begin{equation}
\label{eq:SpatiotemporalPotential_ModeBasis}
    U_C(\mathbf{x},t) = -\sum_{k=1}^{M} u_k(t) \frac{f_k^+(\mathbf{x})}{\phi_0(\mathbf{x})} \,,
\end{equation}
which exploits the orthogonality of the mode functions -- i.e. $2\int d^2\mathbf{x}f_j^-(\mathbf{x})f_k^+(\mathbf{x})=\delta_{jk}$ -- to separately address each collective mode, i.e.
\begin{align}
\label{eq:LinearizedFBPotential}
    \hat{H}_\text{fb}(t) = \sum_j u_j(t) \hat{X}_{j}\,.
\end{align}
Defining the vector of control coefficients, $\zainvec{u}(t) =  \{u_1(t),0,\dots, u_j(t),0,\dots \}^\intercal$, Eq.~\eqref{eq:LinearizedFBPotential} can be written in matrix form as
\begin{subequations}
	\label{eq:ControlHamiltonian_BogBasis}
	\begin{align}
    \hat{H}_\text{fb}    &= -\zainvec{\hat{x}}^\intercal\mm{\Sigma}\mm{B}\zainvec{u}(t)   \,,
\end{align}
\end{subequations}
where the matrix $\mm{B}$ encodes the constraint that the control potential may not be chosen to give a Hamiltonian term proportional to the momentum quadratures $\hat{Y}_j$:
\begin{align}
    \mm{B} = \bigoplus_j^M \begin{pmatrix} 0 && 0 \\ 1 && 0 \end{pmatrix} \,.
\end{align}
Notably this matrix is not of full row rank, nor is it invertible, implying that the feedback scheme we consider cannot implement optimal Markovian control as described in Ref.~\cite{Wiseman2005}.

The addition of the feedback Hamiltonian in Eq.~\eqref{eq:ControlHamiltonian_BogBasis} results in a modified equation of motion for the means~\cite{WisemanMilburn2009}:
\begin{align}
    \label{eq:EOM_Means_WithFeedback}
    d \langle \hat{\zainvec{x}} \rangle &= \mathbf{\Sigma} \mm{A} \langle \hat{\zainvec{x}} \rangle dt + \mm{B}\zainvec{u}(t)dt+ 2\sqrt{ \alpha \eta} \mm{V}(t) \mm{L} d\zainvec{w}(t) \,.
\end{align}
Importantly, the equation of motion for the covariance matrix $\mm{V}$ remains unchanged. This means that the evolution of the covariances depends solely on the measurement, independent of the choice of control, which only needs to be effective in controlling the quadrature means, $\langle \zainvec{\hat{x}} \rangle$. In this sense, the design of an effective control is essentially a classical problem -- though the heating effects of measurement backaction, which is an inherently quantum effect, will define the constraints on the control optimization as we will see later in Section~\ref{sec:SteadyStateAnalytic}.

\section{State estimation for ground-state cooling}
The primary control objective we consider in this work is the ground-state cooling low-energy excitations in a BEC system. Thus, we will instead consider the reduction of energy of each quasi-particle mode as a convenient metric for effective cooling. Specifically, we define the primary objective of the feedback control scheme to be to reduce the (unconditional) mean `phonon' occupation of each quasi-particle mode below unity, i.e. 
\begin{align}
    \bar{n}_j \equiv \expens{\hat{b}_j^\dag \hat{b}_j} \leq 1\,.
\end{align}
In optomechanical and micromechanical systems, this is known as \emph{ground-state cooling}~\cite{Genes2008}. The unconditional phonon occupation can equivalently be expressed in terms of the quadrature operators (\emph{c.f.} Eq.~\eqref{eq:Ham_Bog}), i.e. 
\begin{align}
	&2\bar{n}_j +1 = \expens{\hat{X}_j^2 + \hat{Y}_j^2}\,, \label{eq:PhononOcc_Decomposed} \\ \notag
	&= \underbrace{\text{Cov}\left( \hat{X}_j, \hat{X}_j \right)+\text{Cov}\left( \hat{Y}_j, \hat{Y}_j \right)}_{\rm quantum}+\underbrace{\exps{\expq{\hat{X}_j}^2}+\exps{\expq{\hat{Y}_j}^2}}_{\rm classical}  \,. 
\end{align}
Here we have separated out two contributions to the phonon occupation: first, from the quantum correlations of the system, described by the mode covariances -- that are driven only by the measurement (see Eq.~\eqref{eq:Covariances}) -- and second, from classical fluctuations in the conditional means of the quadrature operators, which are driven by both the feedback and measurement backaction (see Eq.~\eqref{eq:EOM_Means_WithFeedback}).

\subsection{The multi-mode cold-damping control }
Next, we must consider how the information obtained from the measurements are used to inform the controlling potential, Eq.~\eqref{eq:SpatiotemporalPotential_ModeBasis}, in order to achieve efficient cooling of the system. As the unmonitored dynamics of the collective BEC excitations behave as uncoupled quantum harmonic oscillators (see Eq.~\eqref{eq:Ham_Bog}), this suggests the optimal choice of the control coefficients in Eq.~\eqref{eq:ControlHamiltonian_BogBasis} should be proportional to the (conditional) mean of the phase quadratures, i.e. $u_j(t) \propto \langle \hat{Y}_j \rangle$. However, the question remains how $\langle \hat{Y}_j \rangle$ should be estimated from the measurement record Eq.~\eqref{eq:MeasSignal_dy}, which encodes only density information.
 
One approach is to construct a Bayesian estimate of the full quantum state in real time, i.e. a quantum filter, that is continuously updated with the new information obtained from the measurement~\cite{WisemanMilburn2009}. For LQG systems, as we consider here, quantum filtering of this type can be achieved using Kalman filtering methods adapted from classical control~\cite{WisemanMilburn2009,Ma2022}, which enables optimal control; such an approach has recently been applied to demonstrate ground-state cooling of an optically-trapped nanoparticle in a room temperature enviroment~\cite{Magrini2021,Tebbenjohanns2021a}. In the present case of a feedback-controlled BEC, implementing quantum state filtering in real-time is a significant technical challenge, even within the LQG theory. For example, processing the spatially-resolved measurement results and decomposing them into the Bogoliubov basis may pose a technical challenge to efficiently implement, possibly resulting in significant time delays that will, in general, degrade the stability and efficacy of the control protocol. Moreover, the quantum filtering approach requires a reliable model of the underlying dynamics, for which our perturbative LQG model may not be sufficient, as the underlying many-body dynamics of a BEC are in general neither linear nor necessarily quantum Gaussian. \par 

 Here we take a more direct approach, where we will apply feedback proportional to the derivative of the density quadrature of each mode, $u_j(t) \sim \partial_t \langle \hat{X}_j \rangle$, estimated directly from the measurement current Eq.~\eqref{eq:MeasSignal_dy} without necessitating an estimate of the full quantum state. This follows the `cold damping' approach developed in the context of feedback-cooled optomechanical systems~\cite{Genes2008}, where the measurement current is low-pass filtered to remove high-frequency fluctuations such that its time derivative becomes well defined~\footnote{Low-pass filtering the measurement current eliminates arbitrarily high-frequency components of the noise term in Eq.~\eqref{eq:MeasSignal_dy}, such that it becomes smooth (i.e. differentiable) within the bandwidth of the filter.}. This is related to the approach of Ref.~\cite{Yamaguchi2023}, where a similar estimation procedure is considered for the control of momentum fluctuations in a homogeneous BEC system with periodic boundary conditions under a mean-field approximation. 

\subsection{The derivative current}
We will construct the estimate of the velocity ($\partial_t\langle \hat{X}_j\rangle$) of each mode -- henceforth referred to as the `derivative current' -- by first low-pass filtering each element of the measurement current $dy_j$, i.e.
\begin{align}
\label{eq:Lowpassfilter_gen}
	I_j(t) = \int_{-\infty}^\infty g_j(t-s)dy_j(s)\,,
\end{align}
 for some filter kernel $g_j(t)$, and then taking the temporal derivative of $I_j(t)$. Specifically, we consider a simple digital RC filter that can be implemented in real-time, described by the causal filter kernel~\cite{Genes2008}:
\begin{align}
\label{eq:CausalFilter_WithTimeDelay}
	g_j(t) = \Omega_{j}\Theta(t-\epsilon)e^{-\Omega_{j}t} \,,
\end{align}
where $\Omega_{j}$ is the bandwidth of the low-pass filter for the $j$th mode, and the Heaviside step function $\Theta(t)$ ensures the control at time $t$ cannot depend on \emph{future} measurements. We have additionally included a technical time delay of the feedback loop, $\epsilon$ (common to all modes), in order to ensure causality of evolution under Eq.~\eqref{eq:EOM_Means_WithFeedback} -- for any finite value of $\epsilon$, the state of the system at time $t+dt$ should depend only on the measured system observables up to time $t$. In this work we will take this technical time delay to be negligible within the bandwidth of the filter (i.e. $\epsilon^{-1} \gg \max_j\{\Omega_{j}\}$), such that we operate in the Markovian limit $\epsilon\rightarrow 0^+$~\cite{Wiseman1993}.

The properties of the filter kernel, Eq.~\eqref{eq:CausalFilter_WithTimeDelay}, are best assessed in the frequency domain, where it takes the form~\footnote{The factor of $\sqrt{2\pi}$ in Eq.~\eqref{eq:Lowpassfilter_FreqSpace} arises due to the unitary convention of the Fourier transform, and is cancelled out when $g_j(t)$ is convolved with another function.}:
\begin{align}
\label{eq:Lowpassfilter_FreqSpace}
\sqrt{2\pi}\tilde{g}_j(\omega) = \frac{1}{1-i\omega/\Omega_j} = \frac{e^{i \Omega_j^{-1} \omega}}{\sqrt{1+\omega^2/\Omega_j^2}} \,.
\end{align}
This expression demonstrates two features of the low-pass filter: first, frequencies larger than the filter bandwidth, $\omega \gg \Omega_j$ are suppressed by a factor proportional to $\Omega_j/\omega$; and second, the filtered time series for the $j$th mode will have a (mode-dependent) time delay of $\tau_j\equiv\Omega_j^{-1}$ as compared to the original.  Both of these features vanish as we take the bandwidth to be arbitrarily large, and the effect of the filter becomes increasingly negligible. In the limit $\Omega_j \rightarrow \infty$, $I_j(t) \rightarrow dy_j(t)$. Note we have chosen the normalization of $g_j(t)$ such that low-frequency ($\omega \ll \Omega_j$) contributions to $dy_j(t)$ are unaffected by the filtering, i.e. $\int_{-\infty}^\infty g_j(t)=1$.\par 


The derivative current, $\boldsymbol{\mathcal{S}}(t)=(\mathcal{S}_1,0,\mathcal{S}_2,\dots)^\intercal$, is then given by the time derivative of the low-pass filtered measurement current, Eq.~\eqref{eq:Lowpassfilter_gen}, i.e.
\begin{align}
  \mathcal{S}_j(t) &= -\frac{d I_j(t)}{dt} 
= \int_{-\infty}^{\infty} h_j(t-s) dy_j(s) \,, \label{eq:DerivativeCurrent_TemporalDef}  
\end{align}
where in the second line we have defined the effective kernel of the derivative filter, 
\begin{align}
\label{eq:derivativefilter_temporal} 
	h_j(t) \equiv -\partial_t g_j(t) = -\Omega_{j}e^{-\Omega_{j}t}\left(\delta(t) - \Omega_{j}\Theta(t) \right)\,. 
\end{align}
Note we have deliberately chosen the sign of the derivative current to be negative, such that its deterministic component (i.e. the `signal') is proportional to $\partial_t \langle \hat{X}_j \rangle$ upon integrating Eq.~\eqref{eq:DerivativeCurrent_TemporalDef} by parts. Experimentally, the derivative can be constructed from finite-differencing the experimental time series on an interval $\Delta t$ much shorter than the temporal bandwidth of the filter, i.e. $\Delta t \ll \min\{\tau_j\}$. Alternatively, the derivative filter, Eq.~\eqref{eq:DerivativeCurrent_TemporalDef}, can be applied in frequency space to the Fourier transformed measurement current (on a discretized time grid), where the derivative kernel is (\emph{c.f} Eq.~\eqref{eq:Lowpassfilter_FreqSpace}):
\begin{align}
\label{eq:DerivFilter_FreqSpace}
	\tilde{h}_j(\omega) = \frac{i\omega}{1-i\omega/\Omega_j} \,.
\end{align}
This kernel takes the form of a standard derivative high-pass filter, and the ideal derivative limit is recovered in the high-bandwidth limit~\cite{Genes2008} (i.e. $h(t) \rightarrow -\delta'(t)$ as $\Omega_j \rightarrow \infty$), although a finite value of $\Omega_j$ is required in order for the derivative of $dy_j(t)$ to be well defined.

\section{Steady-state ground-state cooling: Analytical Theory \label{sec:SteadyStateAnalytic}}

A powerful advantage of the LQG theory developed above is that it is analytically tractable, and can thus provide broad insights into the key physics of feedback-cooled BECs.  In this section we develop an analytic theory describing the steady-state of BEC collective excitations under continuous measurement and feedback, within the approximation of vanishing measurement-induced couplings between different modes~\cite{Wade2016a}. The aim of this analysis is to identify parameter regimes of effective cooling, focusing primarily on the steady-state phonon occupation. This will lay the groundwork for the calculations performed in Sec.~\ref{sec:Numerics}, where we will demonstrate the feasibility of multi-mode feedback cooling through direct numerical simulations.

\subsection{The decoupled modes approximation \label{sec:DecoupledModesApprox}}

In previous work by Wade \emph{et al}.~\cite{Wade2015,Wade2016a}, a `decoupled modes' approximation was employed to analytically solve a related LQG model of a quasi-1D BEC subject to stroboscopic dispersive measurements. In this approximation scheme, measurement-induced couplings between different quasi-particle modes are neglected, i.e.
\begin{align}
\label{eq:DecoupledModesApprox}
	\mathcal{M}_{jk} &\approx \delta_{jk} \mathcal{M}_{jj} \,,
\end{align}
in which case the matrix $\mm{M}$ becomes block diagonal:
\begin{align}
	\mm{M} &= \bigoplus_j \begin{pmatrix}
		\mathcal{M}_{jj} && 0 \\
		0 && 0 
	\end{pmatrix} \,. 
\end{align}
If we then restrict the control coefficients for each mode, $u_j(t)$, to be a function of only the measurement current for that mode, $dy_j(t)$ -- that is, if we choose $u_j(t)$ to be independent of $dy_k(t)$, for $k \neq j$ -- then the dynamics of each mode described by Eqs.~\eqref{eq:EOM_Means_WithFeedback} and \eqref{eq:Covariances} become completely independent. This significantly reduces the complexity of the system dynamics to that of $M_B$ independent quantum harmonic oscillators, each subject to continuous monitoring and feedback.

\begin{figure*}
	    \centering
    \includegraphics[width=\textwidth]{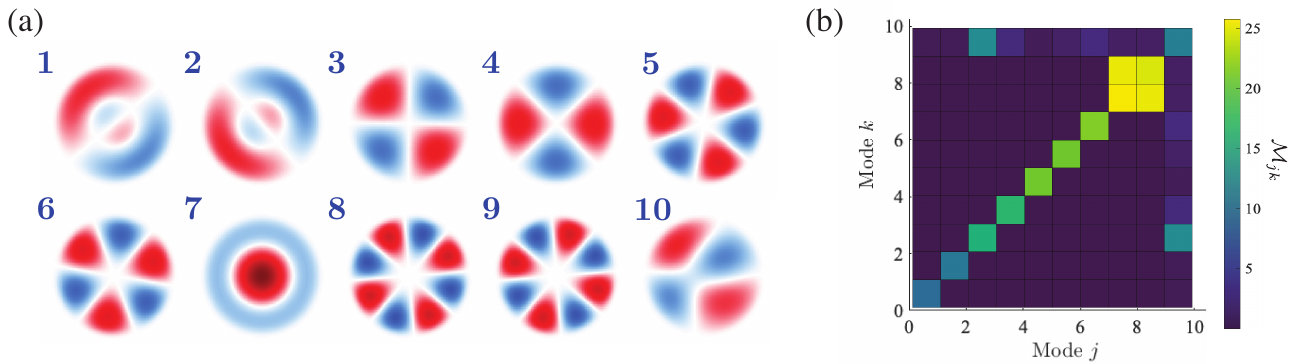}
    \caption{Collective mode structure for the first ten excitations of a quasi-2D cylindrically symmetric BEC, with $N_0=10^5$ condensate atoms and $g_{\rm 2D} = 0.035 E_0/l_0^2$. (a) Mode functions describing density fluctuations, $m_j(x)$, \emph{c.f.} Eq.~\eqref{eq:linearizedMeasurementOperator}. (b) Measurement induced couplings, $\mathcal{M}_{jk}$, for each mode (diagonals) and between modes (off-diagonals). Note that while diagonal terms dominate, there are non-negligible cross-mode couplings between modes $8$ and $9$. }
    \label{fig:modes_diagram}
\end{figure*}

In Fig.~\ref{fig:modes_diagram}, we present the collective mode structure for a cylindrically-symmetric 2D BEC with $N_0=10^5$ condensate atoms, based on numerical diagonalization of Eq.~\eqref{eq:BdGequations} following the methods outlined in Sec.~3.6 of Ref.~\cite{Bisset2013b}. From Fig.~\ref{fig:modes_diagram}(ii), we can see that the off-diagonal elements of $\mathcal{M}_{jk}$ are far weaker than the diagonal elements, providing an empirical justification of the decoupled modes approximation. An alternative motivation for the decoupled modes approximation can be found by considering the analytical solutions to Eq.~\eqref{eq:BdGequations} for quasi-1D systems in the Thomas-Fermi regime~\cite{Petrov2000}:
\begin{subequations}
\label{eq:ThomasFermi1DBogFunctions}
	\begin{align}
	n_0(\tilde{x}) &=\frac{\mu}{g_{\rm 1D}}\left(1-\tilde{x}^2 \right)\,, \\ 
		f_j^-(\tilde{x}) &= \sqrt{\frac{j+1/2}{2R_{\rm TF}}}\left(\frac{2\mu }{\hbar\omega_j}(1-\tilde{x}^2) \right)^{-1/2}P_j(\tilde{x})\,,
\end{align}	
\end{subequations}
where $\tilde{x}=x/R_{\rm TF}$ is the coordinate with respect to the Thomas-Fermi radius $R_{\rm TF} = \sqrt{2\mu/m}$. Note that both of these functions are defined in the range $|\tilde{x}| \leq 1$, and are zero elsewhere. Then, using $m_j(x) = 2\sqrt{n_0}f_j^-(x)$ and $\mathcal{M}_{jk}	=\int dx m_j(x) m_k(x)$, we can analytically compute the couplings:
\begin{align}
	\mathcal{M}_{jk} &= 4R_{\rm TF}\int d\tilde{x} \;n_0(\tilde{x})f_j^-(\tilde{x})f_k^-(\tilde{x})\\ 
	&= \delta_{jk}\frac{\hbar\omega_j}{g_{\rm 1D}}\,,
\end{align}
where in the last line we have used the orthogonality of the Legendre polynomials, i.e. $\int dx P_j(x) P_k(x) =\delta_{jk}/(j+1/2)$. Given that the Thomas-Fermi approximation becomes exact in the large $N_0$ limit, the weak off-diagonal couplings should be understood as finite-size effects. 

\subsection{Quantum correlations and the `weak measurement' regime \label{sec:CovarianceAnalytic_WeakMeasurement}}
The quantum correlations of the continuously-monitored system are captured by the symmetrized covariance matrix $\mm{V}$, which depends only on the strength of the measurement. The steady state of this matrix, denoted as $\mm{V_\infty}$, is straightforwardly obtained by setting $\dot{\mm{V}}=0$ in its dynamical equation of motion, Eq.~\eqref{eq:Covariances}, i.e.
\begin{align}
\label{eq:VssRiccatiEquation}
	0 &= (\mm{\Sigma} \mm{A}) \mm{V_\infty}+ \mm{V_\infty}(\mm{\Sigma} \mm{A})^\intercal + \alpha\mm{D} - 4 \alpha \eta \mm{V_\infty} \mm{M} \mm{V_\infty} \,.
\end{align}
This is an algebraic Riccati equation and may be analytically solved in the decoupled-mode approximation, yielding the block diagonal matrix solution~\cite{Wade2016a}:
\begin{align}
\label{eq:Vss_decoupled} 
\mm{V_\infty} &= \bigoplus_j\frac{1}{4 \eta \tilde{\Gamma}_j}
				\begin{pmatrix}
							\sqrt{2(a_j-1)}	&&	a_j-1 \\
							a_j-1	&&	a_j\sqrt{2(a_j-1)} 
						\end{pmatrix} \,, 
\end{align}
where $\tilde{\Gamma}_j\equiv \Gamma_{j}/\omega_j=\alpha\mathcal{M}_{jj}/\omega_j$ and $a_j = \sqrt{4\eta \tilde{\Gamma}_j^2+1}$. For sufficiently weak measurement, the covariances may be approximated as:
\begin{align}
\label{eq:Vss_weak} \mm{V_\infty} &= \bigoplus_j\frac{1}{2}\begin{pmatrix}
							\eta^{-1/2}&&	\tilde{\Gamma}_j \\
							\tilde{\Gamma}_j	&&	\eta^{-1/2}
						\end{pmatrix}	+ \mathcal{O}\left(\tilde{\Gamma}_j^2 \right)
\end{align}
In the limit of perfect detection, $\eta = 1$, the leading  contribution to $\mm{V_\infty}$ is precisely the covariance matrix describing the ground state of each mode (i.e. the quasi-particle vacuum). Furthermore, the measurement-induced quantum correlations only contribute to the phonon occupation Eq.~\eqref{eq:PhononOcc_Decomposed} at second order in $\tilde{\Gamma}_j	$; it is in this sense that $\tilde{\Gamma}\ll 1$ defines the `weak measurement' regime. Generalizing to the case of finite detection inefficiency, i.e. $0<\eta\leq 1$, the steady-state phonon occupation is bounded by
\begin{align}
\label{eq:MinPhononLimit}
	\bar{n}_j \geq n_{\rm min} \equiv \frac{1}{2}\left( \eta^{-1/2}-1\right) \,.
\end{align}
This bound becomes an equality only if there are no residual classical fluctuations in the quadrature means, which can only be achieved with perfect knowledge of the quantum state (i.e. using a perfectly-convergent quantum state filter). 

\subsection{Spectra of classical fluctuations under feedback}
The goal of the feedback is to eliminate the dynamics of the conditional means of the quadratures, i.e. we will aim to design a control that achieves $\langle \hat{X}_j \rangle = \langle \hat{Y}_j \rangle = 0$ as closely as possible in the steady state. Such a steady-state analysis is most conveniently performed in the frequency domain, which enables us to incorporate the time delay induced by the filter kernel Eq.~\eqref{eq:derivativefilter_temporal}, following the approach developed for optomechanical systems in Ref.~\cite{Genes2008}.\par 

We start with the equations of motion for the conditional means of a mode $j$, using the Stratonovich form of Eq.~\eqref{eq:EOM_Means_WithFeedback}:
\begin{align}
	\dot{X}_j &= \omega_j Y + 2\sqrt{\eta\Gamma_j} V_{j}^{xx} \xi_j(t)\,, \\
	\dot{Y}_j &= (-\omega_j X+u_j(t))+2\sqrt{\eta\Gamma_j} V_{j}^{xy} \xi_j(t) \,,
\end{align}
where $\xi_j(t)$ is the Stratonovich noise corresponding to the Wiener increment $dw_j(t)$, and we have adopted the shorthands: $X_j :=\expq{\hat{X}_j}$, $Y :=\expq{\hat{Y}_j}$, $V_{j}^{xx}:=[V_{j}^{xx}]_{jj}$ and  $V_{j}^{xy}:=[V_{j}^{xy}]_{jj}$. The control term $u_j(t)$ corresponds to the filtered derivative of the measurement current Eq.~\eqref{eq:DerivativeCurrent_TemporalDef}, which can be expressed in the frequency domain as:
\begin{align}
	\tilde{u}_j(\omega) = - G_j(\omega)\left(2\sqrt{\eta\Gamma_j}\tilde{X}_j(\omega)+\tilde{\xi}_j(\omega) \right) \,,
\end{align}
where $\tilde{X}_j(\omega)$ and $\tilde{\xi}_j(\omega)$ denote the Fourier transforms of $X_j(t)$ and $\xi_j(t)$, respectively, and $G_j(\omega)$ is the feedback transfer function (\emph{c.f.} Eq.~\eqref{eq:DerivFilter_FreqSpace})~\cite{Genes2008}: 
\begin{align}
\label{eq:FeedbackTransferFunction}
	G_j(\omega) &= \frac{-i\omega c_j}{1-i\omega/\Omega_j} \,. 
\end{align}
Here we have re-scaled the control gain $k_j \rightarrow c_j \equiv\mathcal{M}_j k_j$; $c_j$ has units of $({\rm time})^{1/2}$, and has the dimensionless form $\tilde{c}_j=c_j\sqrt{\omega_j}$.

\begin{figure*}
	    \centering
\includegraphics[width=0.85\textwidth]{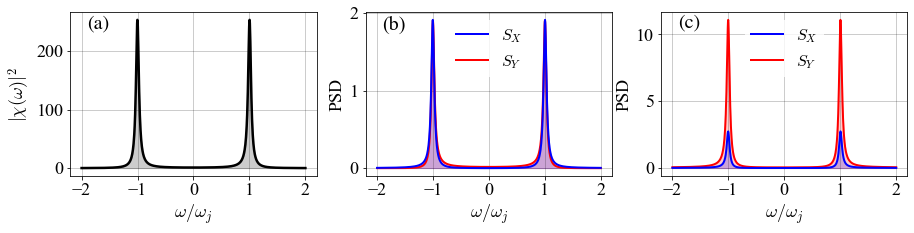}
    \caption{Steady-state fluctuation spectra of a continuously-monitored and feedback-cooled mode with frequency $\omega_j$, in the case of perfect detection ($\eta=1$). (a) Absolute value of the mechanical susceptibility, Eq.~\eqref{eq:MechanicalSusceptDef}, for $\tilde{\Gamma}_j=c_j\sqrt{\omega_j}=\omega_j/\Omega_j = 0.1$ (b-c). (b) Power spectral densities (PSDs) of each quadrature, for the same parameters as (a). (c) Quadrature PSDs for a highly non-thermal steady-state, associated with the parameters $\tilde{\Gamma}_j=0.4$, $\omega_j/\Omega_j=0.35$, and $c_j\sqrt{\omega_j}=0.75$. For both sets of parameters, the shift in the effective resonance frequency due to the control is negligible, i.e. $\omega_j^{\rm eff}\approx \omega_j$.}
    \label{fig:FluctuationSpectra}
\end{figure*}

Using Eq.~\eqref{eq:FeedbackTransferFunction}, we can solve the equations of motion for the means completely in Fourier space: 
\begin{widetext}
\begin{align}
	\tilde{X}_j(\omega) &= \chi_j(\omega)\left(2i\sqrt{\Gamma_j}(V_{j}^{xx}\omega + iV_{j}^{xy}\omega_j) +\omega_j G_j(\omega) \right)\tilde{\xi}_j(\omega), \\
	\tilde{Y}_j(\omega) &= \chi_j(\omega)\bigg(2\sqrt{\Gamma_j}(i V_{j}^{xy}\omega + V_{j}^{xx}\omega_j) -(i\omega-4\Gamma_j V_{j}^{xx})G_j(\omega) \bigg)\tilde{\xi}_j(\omega)  \,.
\end{align}
\end{widetext}
In analogy to the theory of controlled optomechanical systems~\cite{Genes2008}, we have defined the effective mechanical susceptibility of the collective mode:
\begin{align}
\label{eq:MechanicalSusceptDef}
	\chi_j(\omega) \equiv \left[\omega^2-\omega_j^2 - 2\sqrt{\Gamma_j}\omega_j G_j(\omega) \right]^{-1} \,,
\end{align}
with $\Gamma_j$ characterizing the damping rate for the harmonic oscillator. In the analogous case of feedback-cooled optomechanical systems, this would be an intrinsic property of the oscillator due to its coupling with the environment. In contrast, here $\Gamma_j$ is a tunable parameter determined by the coupling of the optical field (i.e. the measurement laser) to low-energy density fluctuations of the atomic cloud. Following Ref.~\cite{Genes2008}, we can identify the (frequency-dependent) cooling rate $\gamma_j(\omega)$ by writing the transfer function explicitly in terms of its real and imaginary components:
\begin{align}
\label{eq:MechanicalSusceptibility}
	\chi_j(\omega) = \left[\omega^2-\omega_j^{\rm eff}(\omega)^2 - i\omega \gamma_j(\omega) \right]^{-1} \,,
\end{align}
where $\omega_j^{\rm eff}(\omega) = \omega_j - 2\sqrt{\Gamma_j} {\rm Re}[G_j(\omega)]\omega_j/\omega$ is the effective oscillator frequency, and 
\begin{align}
\label{eq:DampingRate_freqdep}
	\gamma_j(\omega) & \equiv 2\sqrt{\Gamma_j} {\rm Im}[G_j(\omega)]\omega_j/\omega = \frac{2c_j\sqrt{\Gamma_j}\omega_j}{1+\omega^2/\Omega_j^2}  \,
\end{align}
is a frequency-dependent damping rate. 
The shift to the effective frequency of the oscillator may be neglected in the regime of large filter bandwidth, where the feedback transfer function Eq.~\eqref{eq:FeedbackTransferFunction} is dominated by its imaginary component, i.e. $\left|{\rm Im}[G_j(\omega)]/{\rm Re}[G_j(\omega)] \right|= \omega_j/\Omega_j \ll 1$ for $\Omega_j \gg \omega_j$. This is demonstrated in Fig.~\ref{fig:FluctuationSpectra}, where the shift to the resonance frequency is shown to be insignificant for filter bandwidths as low as $\Omega_j\sim 3\omega_j$.

Next, we compute the power spectral density (PSD) of each quadrature:
\begin{subequations}
\label{eq:PSD_Quad}
	\begin{align}
	S_j^{X}(\omega) &=\frac{1}{2\pi}\int_{-\infty}^\infty d\omega' \exps{\tilde{X}_j(\omega)^*\tilde{X}_j(\omega')} \,, \\
	S_j^{Y}(\omega) &=\frac{1}{2\pi}\int_{-\infty}^\infty d\omega' \exps{\tilde{Y}_j(\omega)^*\tilde{Y}_j(\omega')}\,,
\end{align}
\end{subequations}
which are defined such that $\int d\omega\; S_j^{X}(\omega) = \exps{X_j^2}$, and similarly for $S_j^{Y}(\omega)$. This calculation is simplified by the nature of our measurement noise being uncorrelated in time, $\exps{\tilde{\xi}_j^*(\omega)\tilde{\xi}_j(\omega')}=\delta(\omega-\omega')$, giving the result $S_j^{X,Y} = f^{X,Y}_j(\omega)\chi_j(\omega)|^2/2\pi$, where:
\begin{widetext}
	\begin{align}\label{eq:SXY_spectra}
	f_j^{X}(\omega) &=  4\Gamma_j\omega_j^2\left( (V_{j}^{xx})^2 \frac{\omega^2}{\omega_j^2}  +(V_{j}^{xy})^2 \right) + \frac{c_j^2\omega^2\omega_j^2-4 c_j\sqrt{\Gamma_j}\omega_j\omega^2\left(V_{j}^{xx}+V_{j}^{xy}\frac{\omega_j}{\Omega_{j}}\right)}{1+\omega^2/\Omega_{j}^2} \,, \\
\notag	f_j^{Y}(\omega) &=4\Gamma_j (V_{j}^{xy})^2\omega^2+\frac{c_j^2\omega^4-4c_j\sqrt{\Gamma_j}\omega_j\omega^2(V_{j}^{xx}+V_{j}^{xy}\frac{\omega^2}{\omega_j\Omega_{j}} + 4V_{j}^{xx}V_{j}^{xy}\frac{\Gamma_j}{\omega_j})}{1+\omega^2/\Omega_{j}^2}  +\frac{4(V_{j}^{xx})^2\Gamma_j( \omega_j^2+\omega^2(\frac{\omega_j}{\Omega_{j}}+2c_j\sqrt{\Gamma})^2)}{1+\omega^2/\Omega_{j}^2} \,.
\end{align}
\end{widetext}
An interesting consequence of the continuous measurement and feedback protocol is an asymmetry to the above spectra, which violates the equipartition theorem. The steady state of this system is therefore a non-thermal state, as is well established for cold-damping schemes in optomechanical systems~\cite{Genes2008}. This can be seen in Fig.~\ref{fig:FluctuationSpectra}, which demonstrates that the PSD peak for the momentum quadrature of each mode is both broader and higher amplitude than the monitored position quadrature if the gain ($c_j$) is `too large', as compared to its optimum value (see Eq.~\eqref{eq:QuadFlucSS_Analytic} below).

\subsection{Steady-state phonon occupation \label{sec:SSPhononOcc}}

Integrating the power spectral densities, Eq.~\eqref{eq:SXY_spectra}, over all frequencies yields the steady-state values of $\exps{X_j^2}$ and $\exps{Y_j^2}$, which we can use to find control parameters that minimize the steady-state phonon occupation from Eq.~\eqref{eq:PhononOcc_Decomposed}. Although this optimization is trivial to implement numerically, we can gain analytic insight in the weak-measurement limit, $\tilde{\Gamma}_j\ll 1$, in which the leading-order contribution to the spectra are given by:
\begin{subequations}
\label{eq:Spectra_WeakMeas_Approx}
\begin{align}
	f_j^{X}(\omega) &\approx 4\Gamma_j (V_{j}^{xx})^2 \omega^2  + \frac{c_j^2\omega^2\omega_j^2-4 c_j\sqrt{\Gamma_j}\omega_j\omega^2V_{j}^{xx}}{1+\omega^2/\Omega_{j}^2}\,,\label{eq:SX_WeakMeas_Approx}\\	
	f_j^{Y}(\omega) &\approx \frac{c_j^2\omega^4-4c_j\sqrt{\Gamma_j}\omega_j\omega^2V_{j}^{xx}+4(V_{j}^{xx})^2\Gamma_j \omega_j^2}{1+\omega^2/\Omega_{j}^2} \,,\label{eq:SY_WeakMeas_Approx}
\end{align}
\end{subequations}
noting $V_{j}^{xy}\propto \Tilde{\Gamma}$, from Eq.~\eqref{eq:Vss_weak} -- we will further take the variances of each mode to be given by the diagonal elements of Eq.~\eqref{eq:Vss_weak}.

In order to analytically integrate over the PSDs over all frequencies to obtain the steady-state quadrature variances, we develop an approximation scheme in Appendix~\ref{app:PSD_ApproxScheme} that retains only the leading-order contributions in the small parameters, $\Gamma_j/\omega_j$, $c_j\sqrt{\Gamma_j}/\omega_j$, and $c_j^2/\omega_j$, under the assumption that the control gain satisfies $c_j=\mathcal{O}(\sqrt{\Gamma_j})$. Employing this approximation scheme, we find the leading-order contribution to the quadrature variances for $\tilde{\Gamma}_j\ll 1$:
\begin{subequations}
\label{eq:QuadFlucSS_Analytic}
	\begin{align}
	\exps{Y_j^2} &\approx \exps{X_j^2}+\frac{c_j^2\Omega_{j}}{2(1+\omega_j^2/\Omega_{j}^2) }\,.
\end{align}
\end{subequations}
This expression demonstrates that the equipartition result $\expens{\hat{X}_j^2} = \expens{\hat{Y}_j^2} $ remains violated even in the weak measurement regime, despite being satisfied by the steady-state covariances Eq.~\eqref{eq:Vss_weak}. We attribute this effect to the time delay of the feedback loop, which induces residual classical fluctuations in the momentum quadrature. The magnitude of this grows quadratically with the control gain, i.e. $\exps{X_j^2}-\exps{Y_j^2}\propto c_j^2\Omega_{j}$, and can be minimized by a judicious choice of control gain $c_j$, as we will discuss in the following section.

Using the above result, Eq.~\eqref{eq:QuadFlucSS_Analytic}, and substituting Eq.~\eqref{eq:Vss_weak} into Eq.~\eqref{eq:PhononOcc_Decomposed}, we arrive at the expression for the steady-state phonon occupation in the weak-measurement regime:
\begin{align}
\label{eq:PhononOcc_SteadyState_Analytic}
	\bar{n}_j \approx & \;n_{\rm min} + \frac{1 }{4 c_j \sqrt{\Gamma_j }}\bigg(\frac{2\Gamma_j}{\eta\omega_j}  + \frac{c_j^2\omega_j-2c_j\sqrt{\Gamma_j/\eta}}{1+\omega_j^2/\Omega_{j}^2}\bigg) \notag \\&+ \frac{c_j^2\Omega_{j}}{4(1+\omega_j^2/\Omega_{j}^2) }\,,
\end{align}
where $n_{\rm min}$ is the minimum achievable phonon occupation given by Eq.~\eqref{eq:MinPhononLimit}. This expression is a key result of this work.


\subsection{Optimizing the control: gain and bandwidth \label{sec:ControloptimizationAnalytic}}
Equation~\eqref{eq:PhononOcc_SteadyState_Analytic} demonstrates that there are three `control parameters' that characterize the steady state of the system for each mode: the control gain ($c_j$), the filter bandwidth ($\Omega_{j}$), and the measurement rate ($\Gamma_j$). 
By inspection of Eq.~\eqref{eq:PhononOcc_SteadyState_Analytic}, the steady-state phonon occupation decreases monotonically with $\Gamma_j$ -- we will see in Sec.~\ref{sec:CoolingRate} that the key role of $\Gamma_j$ is to determine the rate at which the controlled system converges to its steady state. Thus, we treat it as a fixed parameter here, and will determine the optimal choices of $c_j$ and $\Omega_{j}$ that minimize $\bar{n}_j$. Unfortunately, analytic minimization of Eq.~\eqref{eq:PhononOcc_SteadyState_Analytic} cannot be done exactly, even in this reduced two-dimensional parameter space.

To tackle this problem, we will optimize the two variables separately, optimizing the control gain first while neglecting terms involving the filter bandwidth, then treating those terms perturbatively. In the first step, we will neglect the time-delay contribution, $c_j^2\Omega_{j} \approx 0$ and further assume the filter bandwidth is sufficiently large, i.e. $\Omega_{j}\gg \omega_j$, such that $[1+\omega_j^2/\Omega_j^2]^{-1}\approx 1$. In this case, the steady-state phonon occupation is approximated by:
\begin{align}
	\bar{n}_j \approx \frac{c_j\omega_j}{4 \sqrt{\Gamma_j } }  + \frac{ \sqrt{\Gamma_j }}{4c\eta\omega_j }  - \frac{1}{2 } \,,
\end{align}
 which is minimized by the following choice of control gain:
\begin{align}
\label{eq:ControlGainOptimal}
	c_j = \sqrt{\frac{\Gamma_j}{\eta \omega_j^2}} \,.
\end{align}
This yields the ideal result, $\bar{n}_j = n_{\rm min}$ in the absence of control noise. This indicates that the filtering of the measurement signal to construct the control is the limiting factor to feedback cooling -- a critical result that has not been taken into account in many previous analyses that have focused on the steady-state fluctuations due to quantum backaction~\cite{Szigeti2010,Hush2013,Wade2016a}. Despite the present approximation scheme neglecting these crucial fluctuations, we will find below that Eq.~\eqref{eq:ControlGainOptimal} provides an accurate estimate of the true optimal control gain (see Fig.~\ref{fig:ParamScan2D_SM}).

Next, we will consider the contribution of the finite bandwidth to the steady-state phonon occupation. Given that the filter bandwidth must necessarily be larger than the mode frequency for an effective control, we may expand Eq.~\eqref{eq:PhononOcc_SteadyState_Analytic} to second order in $(\omega_j/\Omega_{j})$, additionally substituting in Eq.~\eqref{eq:ControlGainOptimal}, giving 
\begin{align}
	\bar{n}_j = n_{\rm min} + \frac{\omega_j^2}{8\sqrt{\eta}\Omega_{j}^2} + \frac{\Gamma_j\Omega_{j}}{4\omega_j^2\eta}  + \mathcal{O}\left[\left(\frac{\omega_j}{\Omega_{j}}\right)^3  \right] \,.
\end{align}
This expression clearly demonstrates the trade-off in choosing the filter bandwidth. As $\Omega_{j}/\omega_j$ becomes larger, the effect of time delay vanishes quadratically (the second term), while the residual noise due to high-frequency fluctuations increases linearly (the third term). This trade-off is minimized by the choice:
\begin{align}
\label{eq:FBBandwidthOptimal}
	\frac{\Omega_j}{\omega_j} = \frac{\eta^{1/6}}{\tilde{\Gamma}_j^{1/3}}  \,,
\end{align}
or equivalently, $\omega_j/\Omega_j = (\tilde{\Gamma}_j/\sqrt{\eta})^{1/3}$. Using this result, and the optimal control gain given by Eq.~\eqref{eq:ControlGainOptimal}, we arrive at the final result for the analytically optimized steady-state phonon occupation from Eq.~\eqref{eq:PhononOcc_SteadyState_Analytic}:
\begin{align}
\label{eq:PhononOcc_SteadyState_Optimal}
	\bar{n}_j = n_{\rm min} + \frac{3\tilde{\Gamma}_j^{2/3}}{8\eta^{5/6}} + \mathcal{O}(\tilde{\Gamma}_j^{4/3}) \,.
\end{align}
Given that this expression monotonically decreases sublinearly with $\tilde{\Gamma}_j$, the measurement rate can always be chosen such that the steady state of the feedback achieves the goal of ground-state cooling, $\bar{n}_j \leq 1$, provided $n_{\rm min} \leq 1$. From Eq.~\eqref{eq:MinPhononLimit} we find that this merely requires $\eta \geq 1/9 \approx 11\%$, which is readily achievable in existing cold-atom experiments.

Together with the optimal control gain Eq.~\eqref{eq:ControlGainOptimal} and filter bandwidth Eq.~\eqref{eq:FBBandwidthOptimal}, Eq.~\eqref{eq:PhononOcc_SteadyState_Optimal} is a central result of this work, demonstrating that the feedback control scheme under present consideration is not only effective in damping the motional dynamics of a BEC, but can asymptotically approach the theoretical limits of cooling,  Eq.~\eqref{eq:MinPhononLimit} -- albeit strictly only in the case of idealized quantum-state estimation. The cost of achieving near-ideal steady-state cooling, by taking $\tilde{\Gamma}_j$ to be asymptotically small, is that the cooling timescale becomes asymptotically large;  we will show in the following section that the rate at which the system converges to its steady state is proportional to $\Gamma_j$. Exploring this trade-off is the focus of the following section.

\subsection{Parameter space for optimal cooling of a single mode \label{sec:ParamScan_SM}}
\begin{figure*}
    \centering
    \includegraphics[width=1.05\textwidth]{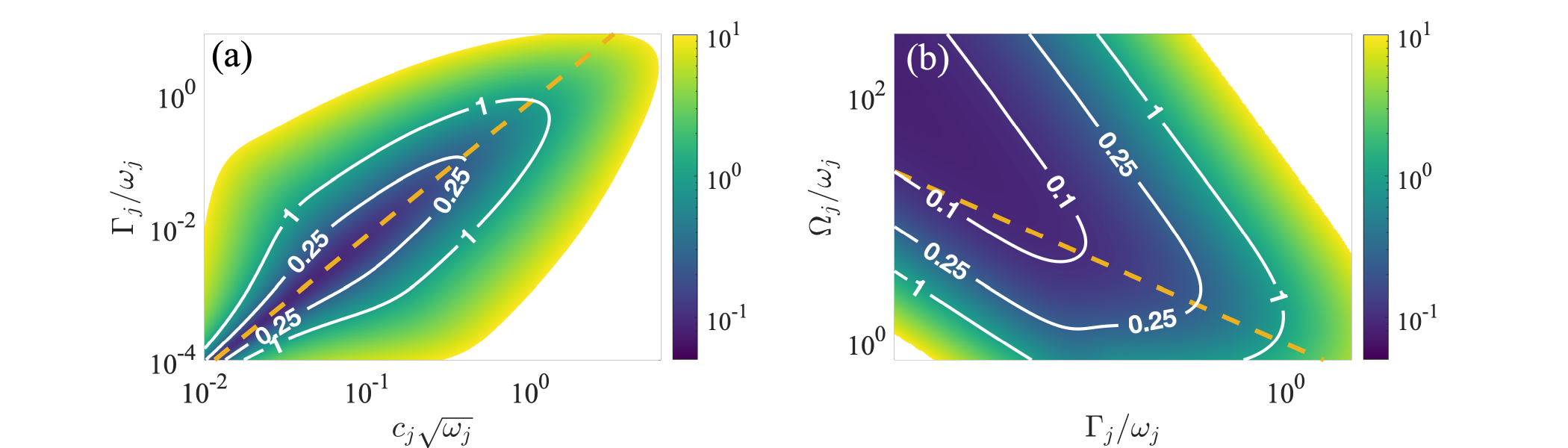}
    \caption{Steady-state phonon occupation of a single mode, $\bar{n}_j$, for a quantum efficiency of $\eta=0.5$ -- using Eq.~\eqref{eq:MinPhononLimit}, this sets the bound $\bar{n}_j\geq\bar{n}_{\rm min}\approx 0.08$. Empty spaces indicate parameter regimes of large steady-state phonon occupations, i.e. $\bar{n}_j>10$. In (a) the filter bandwidth is fixed as a function of $\Gamma_j$, by the analytical result, Eq.~\eqref{eq:FBBandwidthOptimal}. Similarly, in (b) the optimal control gain is chosen using Eq.~\eqref{eq:ControlGainOptimal} for each value of $\Gamma_j$. Dashed (orange) lines indicate the analytical results for (a) the optimal control gain, Eq.~\eqref{eq:ControlGainOptimal}, and (b) the optimal filter bandwidth, Eq.~\eqref{eq:FBBandwidthOptimal}. 
    \label{fig:ParamScan2D_SM}}
\end{figure*}

In Fig.~\ref{fig:ParamScan2D_SM} we explore the dependence of steady-state phonon occupation as a function of control gain $c_j$, measurement rate $\Gamma_j$, and filter bandwidth $\Omega_{j}$. $\bar{n}_j$ is computed numerically from the quadrature spectra~\footnote{To integrate over the quadrature spectra, we used a discrete frequency grid with step size $\Delta \omega/\omega_0 = 10^{-2}$, and restrict the domain to $|\omega|\leq \max \{ 10^3\omega_0,10\Omega_j\}$.}, Eq.~\eqref{eq:SXY_spectra}, i.e.
\begin{align}
\label{eq:PhononOccSteadyState_Computation}
	\bar{n}_j = \frac{1}{2}\int_{-\infty}^\infty d\omega \left(S^X_j(\omega)+S^Y_j(\omega) \right) + \frac{1}{2}\left(V_j^{xx}+V_j^{yy} - 1 \right) \,.
\end{align}
The matrix elements $V_j^{xx}$ and $V_j^{yy}$ in the above expression are computed using the analytic solution for the steady-state covariances in the decoupled-modes approximation, Eq.~\eqref{eq:Vss_decoupled}.
In Fig.~\ref{fig:ParamScan2D_SM}~(a), we find the analytic estimate for the optimal gain coefficient -- Eq.~\eqref{eq:ControlGainOptimal} -- is in good agreement with the numerically-identified optimum values of $c_j$ in the weak measurement regime, $\tilde{\Gamma}_j\lesssim 1$. In this plot, we have taken the filter bandwidth to be given by the analytic result, Eq.~\eqref{eq:FBBandwidthOptimal}, for each value of $\tilde{\Gamma}_j$. This confirms that the optimal control gain should scale with the measurement rate as $c_j\propto\Gamma_j^{1/2}$. The parameter space spanned by the measurement rate and the filter bandwidth (see Fig.~\ref{fig:ParamScan2D_SM}~(b)) shows more interesting behaviours. First, we find the analytic result for the filter bandwidth, Eq.~\eqref{eq:FBBandwidthOptimal}, agrees closely with the numerical optimum in the range $\tilde{\Gamma}_j \sim 10^{-2}-10^{0}$. The analytic result diverges from the numerically-identified optimum for $\tilde{\Gamma}_j\ll 10^{-2}$, which are close to the resonance frequency. This divergence arises due to higher-order contributions in $(\omega_j/\Omega_{j})$ neglected in Eq.~\eqref{eq:FBBandwidthOptimal}. Secondly, Fig.~\ref{fig:ParamScan2D_SM}~(b) demonstrates that the use of Eq.~\eqref{eq:ControlGainOptimal} to determine the control gain fails to realize an effective control outside of the weak-measurement regime -- that is, there is no choice of filter bandwidth that enables ground-state cooling for $\tilde{\Gamma}_j \gtrsim 1$. This is to be expected, as Eq.~\eqref{eq:ControlGainOptimal} is explicitly formulated in the weak measurement regime. 

For very weak measurements in the regime $\tilde{\Gamma}_j \lesssim 10^{-2}$, Fig.~\ref{fig:ParamScan2D_SM}(b) demonstrates the optimal choice of $\Omega_{j}$ is larger than predicted by the analytic result, Eq.~\eqref{eq:FBBandwidthOptimal}. This suggests the possibility of realising ground-state cooling in the fast-feedback regime of $\Omega_{j}\gg \omega_j$, where time delays induced by the feedback loop vanish and the Markovian nature of the system dynamics is recovered, at the cost of significantly reducing the cooling rate (proportional to $\Gamma_j$; see Eq.~\eqref{eq:CoolingRateAnalytic_WeakMeasurement} below). This is appealing as a potential strategy to mitigate the effect of technical time delays (e.g. those induced by electronics and computations in the feedback loop) which will be exacerbated for low bandwidth filters $\Omega_{j}\gtrsim \omega_j$.  Indeed, Fig.~\ref{fig:ParamScan2D_SM}~(b) reveals a broad range of $\{ \Gamma_j, \Omega_{j} \}$ for which cooling to $\bar{n}_j=\mathcal{O}(10^{-1})$ is achievable, suggesting robust schemes can be identified for realistic experimental constraints and technical challenges.


\section{Multi-objective control: steady-state or speed?}

In addition to the steady-state energy, we are interested in the rate at which the system approaches its steady state under feedback, as the ultimate goal is to use feedback cooling as part of the initial-state preparation for quantum gas experiments. In order to realise this goal, the convergence of the control to its steady state must be significantly faster than the typical coherence times of cold-atom systems -- which can be as large as a few seconds for all-optical traps.

\subsection{Analytic estimation of cooling rates \label{sec:CoolingRate}}
In order to determine the rate of cooling, we must consider the rate at which both the quadrature covariances and classical correlations converge to their steady-state values. The calculations presented below are adapted from the single-mode analysis presented in Ref.~\cite{Szigeti2013a}. 

We will begin with the covariance matrix $\mm{V}$, as its evolution is governed by the differential Riccati equation \eqref{eq:Covariances}, which is deterministic, and is independent of the feedback. Defining the difference matrix $\mm{\tilde{V}}\equiv \mm{V}-\mm{V_\infty}$, where $\mm{V_\infty}$ is the steady-state covariance matrix, and using Eq.~\eqref{eq:Covariances} and $\partial_t\mm{V}_\infty=0$, one can write
\begin{align}
    \dot{\tilde{\mm{V}}} = \dot{\mm{V}} &= \mm{\tilde{A}\tilde{V}}+\mm{\tilde{V}\tilde{A}^\intercal} - 2\alpha\eta \mm{\tilde{V}M\tilde{V}}\,,
\end{align}
where $\mm{\tilde{A}}=\mm{\Sigma A} - 2\alpha\eta \mm{V_\infty M}$. This is also a differential Riccati equation, which has a solution of the form
\begin{align}
    \tilde{\mm{V}} &= e^{\mm{\tilde{A}}t}\left(\mm{\tilde{V}_0}-2\alpha\eta\int_0^t dse^{-\mm{\tilde{A}}s}\mm{\tilde{V}M\tilde{V}}e^{\mm{\tilde{A}}s} \right)e^{-\mm{\tilde{A}}t} \,,
\end{align}
with initial condition $\mm{\tilde{V}_0}=\tilde{\mm{V}}(t=0)$. Then, provided there exists a steady-state solution to the Riccati equation Eq.~\eqref{eq:Covariances}, we will have $\mm{\tilde{V}}\rightarrow\mm{0}$ in the large time limit. In this limit, we may approximate the above equation as
\begin{align}
    \tilde{\mm{V}} &\approx e^{\mm{\tilde{A}}t}\mm{\tilde{V}_0}e^{-\mm{\tilde{A}}t} \,.
\end{align}
As $\mm{\tilde{A}}$ must be negative-definite in order for a steady state to exist, the right-hand side of the above equation is bounded by the real part of the slowest decaying eigenvalue of $\mm{\tilde{A}}$, i.e.
\begin{align}
\label{eq:BoundCovariancesConvergence}
    \tilde{\mm{V}} &\approx e^{\mm{\tilde{A}}t}\mm{C}e^{-\mm{\tilde{A}}t} \leq \exp\left(-2 \;\text{max}\;\textrm{Re}\left\{\bm{\lambda}[\mm{\tilde{A}}]\right\}t\right) \,,
\end{align}
where $\bm{\lambda}[\mm{G}]$ notates the eigenvalues of the matrix $\mm{G}$. Therefore we may take the rate of convergence of the covariance matrix to steady state to be given by the slowest decaying eigenvalue: $r^\mm{V} = 2 \;\text{max}\left(\textrm{Re}\left\{ \bm{\lambda}[\mm{\tilde{A}}]\right\}\right) $. Using the weak-measurement result for the covariances, Eq.~\eqref{eq:Vss_weak}, we find that the quantum correlations of the $j$th mode exponentially converge to their steady-state values at the rate:
\begin{align}
	r^\mm{V}_j = \sqrt{\eta}\Gamma_j \,.
\end{align}

We may follow a similar approach to determine the rate of convergence for the conditional quadrature means in in terms of the correlation matrix 
\begin{align}
	\mm{H}(t)\equiv\exps{\expq{\zainvec{\hat{x}}}(t)\expq{\zainvec{\hat{x}}^\intercal}(t) } \,.
\end{align}
This matrix encodes classical correlations of the system, which the feedback control aims to eliminate (\emph{c.f.} Eq.~\eqref{eq:PhononOcc_Decomposed}). The rate of cooling is thus determined by the rate at which $\mm{H}$ converges to its steady state.
Although an approximate Riccati equation describing the evolution of this matrix may be derived, doing so is highly involved due to the non-Markovian effects induced by the filtering -- specifically, non-negligible temporal correlations in the filtered measurement current, Eq.~\eqref{eq:DerivativeCurrent_TemporalDef}. Furthermore, a complete dynamical description of $\mm{H}$ is unnecessary for this analysis, which requires only a description of the deterministic evolution of the quadrature means (i.e. the `drift' term in the Riccati equation), described by (\emph{c.f.} Eq.~\eqref{eq:EOM_Means_WithFeedback}):
\begin{align}
\label{eq:MeansConvergence_NoNoise}
     \frac{d \langle \hat{\zainvec{x}} \rangle}{dt} &= \bigoplus_j \begin{pmatrix}
 		0 & \omega_j \\
 		-\omega_j & -2 c_j \sqrt{\Gamma_j \eta}\omega_j
 	\end{pmatrix}\langle \hat{\zainvec{x}}\rangle\equiv \mm{G} \langle \hat{\zainvec{x}}\rangle \,,
\end{align} 
where we have defined $\mm{G}$ as the block-diagonal matrix given in the first equality. We then assume the equation of motion for the correlation matrix takes the form of a differential Riccati equation, i.e.
\begin{align}
	\label{eq:RiccatiEq_H} \dot{\mm{H}} &= \mm{G}\mm{H} + \mm{H}\mm{G}^\intercal +\mm{Q} \,,
\end{align}
where we have included a general diffusion term, $\mm{Q}$, which represents the net effect of all noises -- that is, both backaction and control noise -- driving the conditional means (see Eq.~\eqref{eq:EOM_Means_WithFeedback}). We will assume the covariances are in their steady state for this calculation, i.e. $\mm{V}(t)\rightarrow\mm{V}_\infty$ -- this is a prerequisite condition for the correlations between the quadrature means to approach a steady state (see Eq.~\eqref{eq:EOM_Means_WithFeedback}).  We need not make any particular assumption regarding the form of $\mm{Q}$ in order to extract the convergence rate as $r^\mm{H} = 2 \;\text{max}\left(\textrm{Re}\left\{ \bm{\lambda}[\mm{G}]\right\}\right)$, following the same argument as for the covariance matrix. From Eq.~\eqref{eq:MeansConvergence_NoNoise}, the eigenvalues of $\mm{G}$ for the $j$th mode are $\omega_j(-c_j\sqrt{\Gamma_j\eta} \pm \sqrt{c_j^2\eta\Gamma_j -1})$, from which we have:
\begin{align}
\label{eq:ConvergenceCorrelations_NoDelay}
	r_j^\mm{H}  =2\left|\textrm{Re}[\omega_j(-c_j\sqrt{\Gamma_j\eta} + \sqrt{c_j^2\eta\Gamma_j -1})] \right| = 2 \omega_j\sqrt{c_j^2\Gamma_j\eta} \,,
\end{align}
assuming $\eta c_j^2\Gamma_j <1$.  Then, substituting the optimal control gain -- Eq.~\eqref{eq:ControlGainOptimal} -- we obtain $r_j^\mm{H}=2\Gamma_j$. Note that the system does not converge to a steady state if the gain is chosen to be too large, i.e. $c_j^2>1/(\eta\Gamma_j)$.

We note the above result assumes the convergence is not significantly affected by time delays due to the finite bandwidth of the control. We can account for this by replacing the control gain $c_j$ with the gain of the feedback transfer function, given by the imaginary component of Eq.~\eqref{eq:FeedbackTransferFunction} evaluated on resonance, i.e.
\begin{align}
	c_j\rightarrow \frac{{\rm Im}[G_j(\omega_j)]}{\omega_j} = \frac{c_j}{1+\omega_j^2/\Omega_{j}^2}  \,.
\end{align}
Substituting this into Eq.~\eqref{eq:ConvergenceCorrelations_NoDelay}, we then arrive at the general expression for the convergence rate of the control:
\begin{align}
\label{eq:ConvergenceRate_MeansCorrelations}
	r_j^\mm{H}  =\frac{2 \omega_j\sqrt{c_j^2\Gamma_j\eta}}{1+\omega_j^2/\Omega_{j}^2} \,.
\end{align}
Then, using the optimal bandwidth given by Eq.~\eqref{eq:FBBandwidthOptimal}, we find
 the convergence rate for the classical correlations of the $j$th mode to be:
\begin{align}
\label{eq:CoolingRateAnalytic_WeakMeasurement}
	r^\mm{H}_j &= \frac{2\Gamma_j}{1+(\tilde{\Gamma}_j/\sqrt{\eta})^{2/3}} \approx 2\Gamma_j\left(1-\left(\frac{\Gamma_j}{\sqrt{\eta}\omega_j}\right)^{2/3}  \right) \,.
\end{align}
This result demonstrates that in the weak measurement regime $\tilde{\Gamma}_j\ll 1$ the cooling rate is not affected by the time delay associated with an optimally-chosen filter bandwidth, to leading order in $\tilde{\Gamma}_j$. This further solidifies the feasibility of feedback cooling the motion of a BEC without necessitating real-time quantum-state estimation.

We note that $\mm{H}$ will not reach its steady state until the covariance matrix has converged to its own steady state, as the equations of motion for the means, Eq.~\eqref{eq:eoms_expectations}, explicitly depend on $\mm{V}(t)$. Nevertheless, we can always consider the covariances to be in their steady state when analysing the efficacy of the feedback control, as we can choose to do a period of measurement with no feedback to establish the steady-state quantum correlations prior to the initialization of the feedback control. In any case, feedback-less measurement must be done for a minimum duration of a few delay periods (with $\tau_j=\Omega_{j}^{-1}$ the mode-dependent delay time), in order to realize the low-pass filtering of the measurement current.

\subsection{Multi-objective optimization: Exploring the trade-off between cooling rate and steady-state energy \label{sec:MultiObjectiveOptimisation}}

To determine the optimal control parameters, we numerically optimize the control parameters within the decoupled-modes approximation in the parameter space spanned by the vector of control parameters $\zainvec{a}=\left(\alpha,c_1,\Omega_1,c_2,\Omega_2,\dots \right)^\intercal$. Although our primary goal here is to reduce the steady-state energy of the system, we must also include some cost to the cooling timescale, $t_C$, to avoid control solutions with arbitrarily slow convergence. This can be achieved by defining a multi-objective cost function that contains weighted contributions from the \emph{total} steady-state energy of the controlled modes,
\begin{align}
\label{eq:Ess_DecoupledModes}
	E_{\infty} = \hbar\sum_{j=1}^M \omega_j\left(\bar{n}_j + \frac{1}{2} \right) \,,
\end{align}
and the timescale of the cooling, $t_C$. The steady-state phonon occupation for each mode is computed within the decoupled-modes approximation from Eq.~\eqref{eq:PhononOcc_Decomposed}, integrating over the quadrature PSDs given by Eq.~\eqref{eq:PSD_Quad} to obtain the steady-state classical correlations of the quadrature means (as in Sec.~\ref{sec:SSPhononOcc}).

For a single mode, we may estimate the cooling timescale from the convergence of correlations between the conditional quadrature means, i.e. $t_{\rm C} \approx 1/r^\mm{H}$. Generalizing to the multi-mode case, we will estimate the cooling timescale based on the slowest of these convergence rates, i.e. 
\begin{align}
\label{eq:CoolingTimescaleMultiMode}
	t_C = \max_j\left[\frac{1+\omega_j^2/\Omega_j^2 }{2\omega_j\sqrt{c_j^2 \alpha\mathcal{M}_j  } }\right] \,,
\end{align}
using $\Gamma_j = \alpha\mathcal{M}_j$. The total cost function describing our optimization problem may then be written as:
\begin{align}
\label{eq:MultiobjectiveCostFunction_NoLoss}
	J[\mathbf{a},w] = w E_{\infty}[\mathbf{a}]  + (1-w)t_C[\mathbf{a}] \,,
\end{align}
where $w\in [0,1]$ is the \emph{weight} that determines the importance of the steady-state energy against the cooling timescale. This defines a family of optimal solutions parameterized by $w$, $\mathbf{a}_{\rm opt}(w)$, i.e.
\begin{align}
\label{eq:MultiObjectiveOptimisation_OptParamsDef}
	\min_{\mathbf{a}} J[\mathbf{a},w]  = J[\mathbf{a}_{\rm opt}(w),w] \,.
\end{align}
\begin{figure}
    \centering
    \includegraphics[width=0.95\columnwidth]{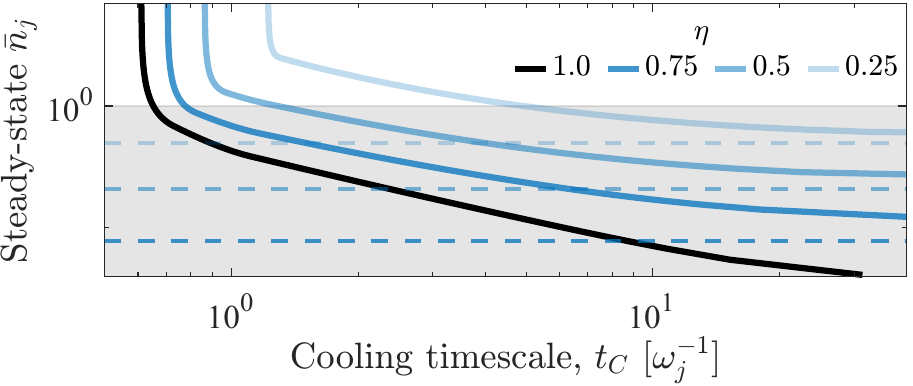}
    \caption{
    Pareto front characterizing the family of optimal solutions to the cost function, Eq.~\eqref{eq:MultiobjectiveCostFunction_NoLoss}, for a single mode of frequency $\omega_j$. Each point along the Pareto front represents the minimum of Eq.~\eqref{eq:MultiobjectiveCostFunction_NoLoss} for a particular weighting, $w$, for both perfect detection (black line) and imperfect detection (blue lines, decreasing opacity with $\eta$). For each value of $\eta<1$, the horizontal dashed lines give the theoretical minimum achievable phonon occupation $\bar{n}_{\rm min}$, given by Eq.~\eqref{eq:MinPhononLimit}. For $\eta=1$, $\bar{n}_{\rm min}=0$. The shaded region indicates $\bar{n}\leq 1$, i.e. the regime of ground-state cooling.
    }
    \label{fig:ParetoFront_SM}
\end{figure}
In Fig.~\ref{fig:ParetoFront_SM}, we present the \emph{Pareto front} representing the family of optimal solutions for the case of a single mode, which qualitatively demonstrates the tradeoff between fast cooling and low steady-state energy for different values of $\eta$. We find that, for modest values of $\eta\gtrsim 0.2$, parameters can be identified that achieve steady-state phonon occupations below unity (i.e. ground-state cooling) on timescales of $\mathcal{O}(1)$ trap periods. Interestingly, Fig.~\ref{fig:ParetoFront_SM} demonstrates that while lower steady-state energies can be achieved at the cost of slower cooling, there are diminishing returns beyond $\omega_j t_C \approx 10$, asymptotically approaching the theoretical limit, Eq.~\eqref{eq:MinPhononLimit}, for large $t_C$.

\section{Dynamical simulations of multi-mode cooling \label{sec:Numerics}}

To conclude this manuscript's investigation, we present numerical simulations of the multi-mode system dynamics within the LQG theory developed in Sec.~\ref{sec:LGTheory}. This is necessary to validate the decoupled-modes approximation, developed in Sec.~\ref{sec:DecoupledModesApprox}, by quantifying the contribution of measurement-induced cross-mode couplings to the steady state of the system. Furthermore, numerical simulations allow us to go beyond the intuitive steady-state analysis to model the transient dynamics of the controlled system, which is required to establish the stability and efficacy of the control.

\begin{figure*}[t!]
    \centering
    \includegraphics[width=1.8\columnwidth]{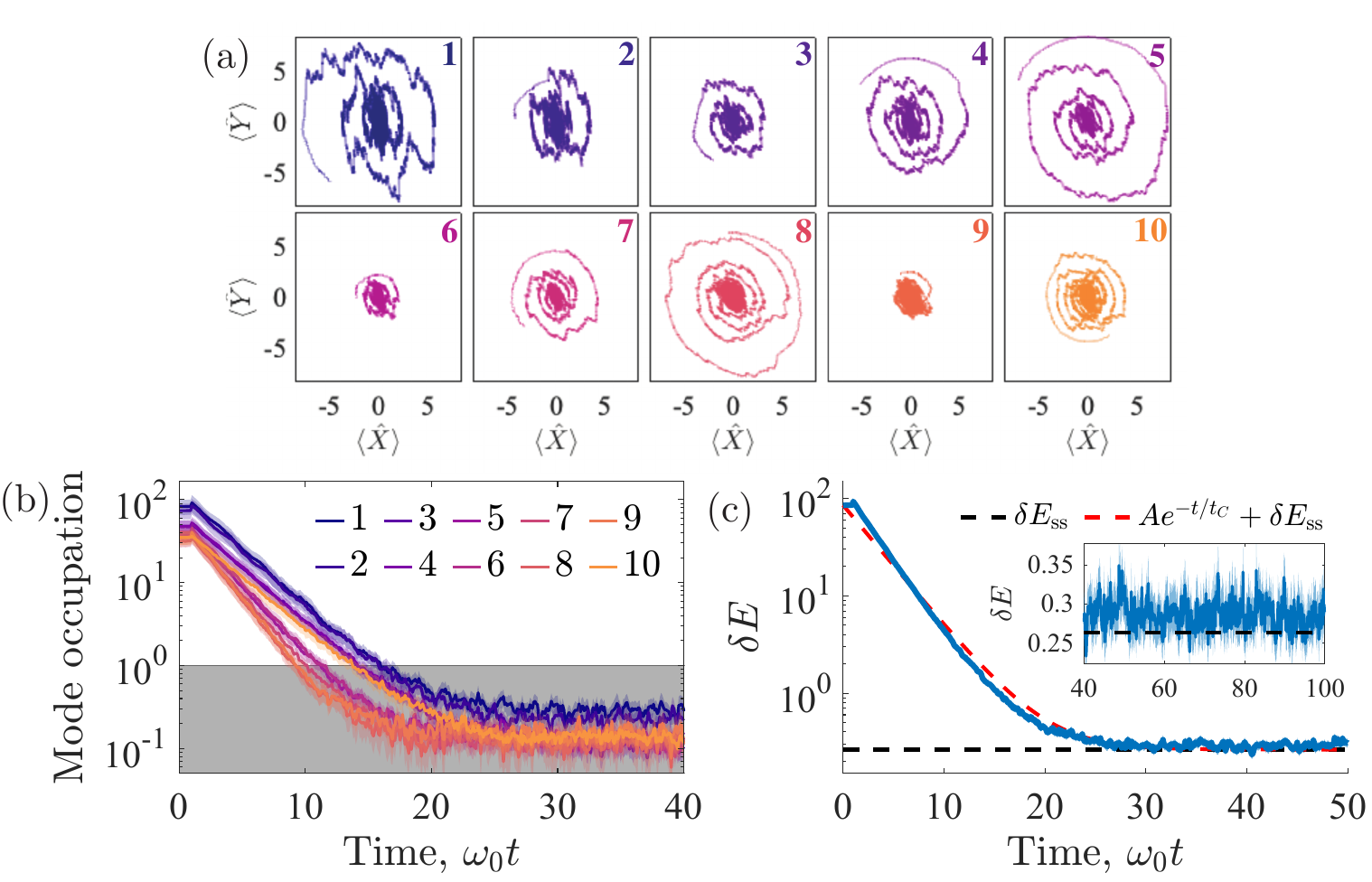}
    \caption[Numerical simulation of feedback cooling, for the lowest $10$ modes of a quasi-2D BEC.]{Numerical simulation of feedback cooling, for the lowest $10$ modes of a quasi-2D BEC (parameters in the main text). (a) Phase-space trajectories of the conditional quadrature means, for a single numerical experiment (i.e. a single measurement record). (b) Unconditional evolution of the phonon occupations of each mode, $\bar{n}_j(t)$, averaged over an ensemble of $100$ numerical experiments -- the shaded region denotes a $95\%$ confidence interval for each data set, estimated as twice the standard error in the mean. The regime of ground-state cooling, $\bar{n}_j \leq 1$, is shaded grey. (c) Relative energy deviation from the zero-point energy $\epsilon_0$, i.e. $\delta E=(E-\epsilon_0)/\epsilon_0$, with long-time evolution within the system's steady state shown in the inset. Blue shading denotes a $95\%$ confidence interval\footnote{The confidence interval in the total energy is estimated by combining the $2\sigma$ error of each mode in quadrature, where $\sigma$ is the standard error in the ensemble average over $100$ `numerical experiments'. This assumes the noise in each mode is independent, which is approximately satisfied within the decoupled modes approximation.}, though this is only visible in the inset. Black and red dashed lines display the semi-analytical predictions for the steady-state energy, Eq.~\eqref{eq:Ess_DecoupledModes}, and cooling timescale, Eq.~\eqref{eq:CoolingTimescaleMultiMode}, respectively. 
    \label{fig:10Modes2DCooling_Numerics}}
\end{figure*}
\subsection{Numerical methods and parameter values}
In this section, we present numerical simulations of the lowest $10$ collective excitations for the quasi-2D BEC parameters considered in Fig.~\ref{fig:modes_diagram}: $N_0=10^5$, $g_{\rm 2D} = 0.035 E_0/l_0^2$, and $\sigma_0 = 0.05 l_0^2$. For the optimization of the measurement and control parameters, we will use \textsc{Matlab}'s constrained local optimization tool, \textsc{fmincon}, to minimize the cost function  Eq.~\eqref{eq:MultiobjectiveCostFunction_NoLoss}. Our target will be to cool to below the ground-state criterion, $\bar{n}_j\leq 1$ for each $j$, within a few tens of trap periods -- cooling on significantly longer timescales cannot be described by numerical simulations of our LQG model, which is inherently perturbative. We will restrict ourselves to the case of perfect detection, i.e. $\eta=1$, which will bound the efficacy of feedback cooling for this system. 

The system dynamics will be modelled under the assumption that the mode covariances begin in their steady state (justified previously in Sec.~\ref{sec:CoolingRate}). The steady-state solution for the covariance matrix, $\mm{V_\infty}$, is numerically found as the stabilizing solution to Eq.~\eqref{eq:VssRiccatiEquation}, using the \textsc{icare} package in \textsc{Matlab}. In this case, the system evolution is described purely by the equation of motion for the conditional quadrature means, Eq.~\eqref{eq:EOM_Means_WithFeedback}, which we numerically integrate using a simple first-order Euler algorithm, with dimensionless timesteps in the range $\omega_0 \Delta t = 10^{-4}-10^{-3}$. Initial conditions of the conditional quadrature means -- i.e. $\langle\hat{X}_j\rangle(t=0),\,\langle\hat{Y}_j\rangle(t=0)$ -- are sampled such that the classical contribution to the phonon occupation, $\bar{n}^{\rm classical}_j= (\exps{\expq{\hat{X}_j}^2}+\exps{\expq{\hat{Y}_j}^2})/2$ (\emph{c.f.} Eq.~\eqref{eq:PhononOcc_Decomposed}), follows the Bose-Einstein distribution:
\begin{align}
	\bar{n}^{\rm classical}_j(t=0) = \frac{1}{e^{\hbar\omega_j/(k_BT)}-1} \,.
\end{align}
In the simulations presented below, we will consider a relatively high-temperature initial condition, $k_B T\approx 72.9 \hbar\omega_0$ -- for $\omega_0/(2\pi)=20$Hz, this corresponds to $T=70$nK and $\bar{n}^{\rm classical}_j(t=0)=\mathcal{O}(10^2)$. 

Although our model of the feedback-cooled BEC assumes a continuous measurement scheme, in practice such a scheme will be realized by stroboscopic measurements at a rate faster than the low-energy dynamics of the atomic cloud. Thus, we will treat the discretized timestep of our numerical simulations, $\Delta t$, as the duration of a single non-destructive measurement. This allows us to treat our simulations as numerical experiments, wherein we explicitly conduct the signal filtering on a discrete time series of measurement results, separated by $\Delta t$.

At each timestep of the simulation, the measurement noise $d\zainvec{w}(t)$ is sampled from a normal distribution with variance $\Delta t$, i.e. as a Wiener increment. This noise not only appears as a driving term in the evolution of the conditional means, Eq.~\eqref{eq:EOM_Means_WithFeedback}, but also in the multi-mode measurement current, Eq.~\eqref{eq:MeasSignal_dy}, which is constructed at every timestep. We explicitly model the filtering of this time-discretized measurement current to construct the control, following a procedure that closely mimics its experimental implementation: for each mode $j$, the low-pass filter kernel (Eq.~\eqref{eq:CausalFilter_WithTimeDelay}) is represented on the discretized time grid on a temporal window of width $3\tau_j$. That is, the exponential tail of $g_j(t)$ is truncated, such that the filtering of the measurement signal (Eq.~\eqref{eq:DerivativeCurrent_TemporalDef}) becomes a weighted sum over $3\,\left\lceil\max_j \{ \tau_j\}/\Delta t\right\rceil$ points in the time series of $dy_j(t)$~\footnote{The truncation of $g_j(t)$ reduces its normalization from unity. To prevent this from affecting the control gain, we re-normalize the filter kernel after it is represented on the discretized
time grid, such that $\sum_{t_n\in [t-3\tau_j,t]} g_j(t_n) = 1$. }. Furthermore, the time derivative of the measurement signal is approximated by a first-order finite difference between neighbouring timesteps to obtain the derivative current, Eq.~\eqref{eq:DerivativeCurrent_TemporalDef}, which is then re-scaled to construct the control coefficient, $u_j(t)$, for the \emph{following} timestep (to ensure causality of the dynamics). Taking $\{t_n\}$ to represent the discretised temporal grid, this procedure can be summarised by the expression:
\begin{align}
	u_j(t_{n+1}) = -k_j \sum_{t_m\in [t_n - 3\tau_j]} g_j(t_n-t_m)\frac{dy(t_m)-dy(t_{m-1})}{\Delta t} \,,
\end{align}
recalling that $k_j=c_j/\mathcal{M}_j$, where $c_j$ are corresponding optimisation variables (\emph{c.f.} Eq.~\eqref{eq:MultiobjectiveCostFunction_NoLoss}).

To find optimal control coefficients using the multi-parameter optimization approach described above (see Eq.~\eqref{eq:MultiObjectiveOptimisation_OptParamsDef}), we will take the weighting in Eq.~\eqref{eq:MultiobjectiveCostFunction_NoLoss} to be $w=0.7$, thereby placing significant (yet secondary) importance on the fast cooling while ensuring all modes are cooled to the ground-state regime, $\bar{n}_j\leq 1$. For the $10$-mode system under consideration, we find the corresponding optimal control parameters to be: $\alpha\approx0.012$, $c_j\sqrt{\omega_j}\approx 0.5-0.2$ (roughly decreasing in magnitude with mode frequency), and $\Omega_j/\omega_j \approx 3-4$ (roughly increasing with mode frequency).

\subsection{Exemplary cooling demonstration}
In Fig.~\ref{fig:10Modes2DCooling_Numerics}, we present the conditional dynamics of the means for a single `numerical experiment' (Fig.~\ref{fig:10Modes2DCooling_Numerics}(a)) -- that is, for a single measurement record characterized by the time series of random numbers, $d\mathbf{w}(t)$ -- as well as the \emph{unconditional} phonon occupation, computed from Eq.~\eqref{eq:PhononOcc_Decomposed} by ensemble averaging over $100$ numerical experiments. Figure~\ref{fig:10Modes2DCooling_Numerics}(a) demonstrates that, for a single experimental run, the conditional means of the quadrature operators for each mode converge towards zero. This suggests the feedback-cooling scheme considered in this work is effective in stabilizing the multi-mode motion of a BEC in real time, despite non-negligible time delays induced by relatively low filter bandwidths.

The unconditional dynamics of the system, shown in Fig.~\ref{fig:10Modes2DCooling_Numerics}(b) and (c), demonstrates feedback cooling of the system close to its ground-state energy, with the steady-state mode occupations for each mode well within the ground-state regime (Fig.~\ref{fig:10Modes2DCooling_Numerics}(b)). In the transient dynamics of the system, the energy decays exponentially (Fig.~\ref{fig:10Modes2DCooling_Numerics}(c)), with the decay timescale in excellent agreement with the semi-analytic estimate of the cooling timescale, Eq.~\eqref{eq:CoolingTimescaleMultiMode}, which gives $\omega_0t_C \approx 3.5$ for the parameters of Fig.~\ref{fig:10Modes2DCooling_Numerics}. Figure~\ref{fig:10Modes2DCooling_Numerics}(c) also shows that the total energy of the system converges to within $30\%$ of the zero-point energy of the system, in excellent agreement with the semi-analytic prediction, Eq.~\eqref{eq:Ess_DecoupledModes} (computed using Eq.~\eqref{eq:PhononOccSteadyState_Computation}). The strong agreement between the numerical and semi-analytic results supports the use of the decoupled-modes approximation for studying and optimizing the feedback control of BEC motion, and validates the analytical results for the steady-state mode occupation and cooling rates derived in Sec.~\ref{sec:SteadyStateAnalytic}.

\section{Summary and Outlook}
In this work, we have theoretically investigated the feedback cooling of quantum-degenerate Bose gases, in the near-equilibrium regime where low-energy collective excitations dominate over particle-like excitations in their contributions to the quantum-statistical properties of the atomic ensemble. We developed an analytically-tractable LQG theory of feedback-cooled BECs, and applied it to study the viability of cooling a quasi-2D BEC to its multi-mode motional ground state without necessitating real-time estimation of the full multi-mode quantum state. Our findings revealed a broad parameter landscape in which simultaneous cooling of many low-energy BEC modes is achievable, which is well captured by straightforward analytical and numerical calculations. The analytic theory developed in this work provides insight into the multi-mode open-quantum system dynamics of feedback-controlled BECs, and is well suited to guide future experimental design, development and analysis.

Although we have focused on the case of cylindrically-symmetric harmonic trapping potentials, the LQG theory developed in this work can be straightfowardly applied to more general traps; the different spatial structure of quasiparticle excitations in different trapping potentials is encoded in the couplings $\mathcal{M}_{jk}$, and do not appear elsewhere in our theory. Therefore, it would be interesting to apply the LQG theory developed here to study the feedback control of BECs in other quasi-2D systems, such as optical lattices, or toroidal BECs. The latter may play an important role in emerging quantum technologies based on `atomtronic' circuits, where measurement-based feedback control may provide avenues for enhanced coherence and robustness of atomtronic devices against motional heating and dephasing.

Multi-mode feedback control over the motional degrees of freedom of ultracold atomic gases also offers interesting prospects for quantum sensing applications, where open-loop quantum control techniques are already being applied~\cite{Gaaloul2022,Saywell2023a,Rodzinka:2024}. A concrete example could be the application of the feedback scheme considered in this work to BEC interferometry experiments in the Cold Atom Lab aboard the International Space Station~\cite{Aveline2020,Gaaloul2022}, for which enhanced motional stabilization provided by closed-loop feedback promises to directly translate into gains in precision and sensitivity. In a similar vein, linear quantum feedback could also be applied to Earth-based atom inteferometers, where motional stabilization of the initial atomic sample could mitigate the effect of spurious inertial accelerations and rotations~\cite{dArmagnacdeCastanet2024,Ben-Aicha2024}. A closely related application is the generation of metrologically-useful entanglement in large cold-atom ensembles using spin-spin interactions and quantum-non-demolition (non-destructive) measurements~\cite{Szigeti2021a}, the realization of which requires ultrastable BEC initial states with carefully controlled initial conditions~\cite{Haine2014,Szigeti2020a,Corgier2021}. In principle, closed-loop feedback could be employed in these systems to eliminate unwanted motion of the initial atomic sample.

An important avenue for future theoretical investigations will be to apply the insights gained in this work to study the prospect of feedback-cooled condensation -- that is, the creation of a BEC from a thermal cloud purely through control of its motional degrees of freedom~\cite{Mehdi2024a}. This will require non-perturbative treatments of the multi-mode atomic dynamics under quasi-continuous measurement and feedback in order to correctly capture the critical fluctuations of the system near the critical temperature of condensation, $T_c$. We have recently developed numerical tools capable of handling the high-dimensional quantum field dynamics based on phase-space representations of the \emph{unconditional} atomic dynamics under measurement and feedback~\cite{Zhu2025}. Although preliminary simulations of a quasi-1D BEC indicate the viability of feedback-cooled condensation, detailed theoretical modeling will also need to incorporate decoherence and heating due to spontaneous emission, which is expected to be the dominant heating source in the thermal regime~\cite{Mehdi2024a}. An interesting aspect of this research will be to compare the multi-mode control scheme developed in this work, where low-lying collective modes are independently controlled, to existing proposals which consider an `energy damping' control that addresses density fluctuations within the bandwidth of the measurement and optical control with spatially-homogeneous gain~\cite{Goh2022a,Mehdi2024a,Zhu2025}.

\section*{Acknowledgments}
The authors acknowledge insightful conversations with Simon A. Haine, and Ryan J. Thomas. SSS was supported by an Australian Research
Council Discovery Early Career Researcher Award
(DECRA), Project No. DE200100495. MLG acknowledges the Rhodes Trust for the support of a Rhodes Scholarship. MLG was also (partially) supported by the Engineering and Physical Sciences Research Council under EPSRC project EP/Y004655/1. MJB was supported by a Gates Cambridge Scholarship (\#OPP1144).

\bibliographystyle{bibsty}
\bibliography{fbc_bib}

\pagebreak
\appendix
\widetext
\section{Derivation of LQG Theory \label{app:LQG_Derivation}}
Here we elaborate on the derivation of the LQG model, Eq.~\eqref{eq:GaussianEvolution}, from the SME Eq.~\eqref{eq:SME_2D} after the approximate diagonalization of the atomic Hamiltonian in Eq.~\eqref{eq:Ham_Bog_quasiparticles} and linearization of the measurement operator in Eq.~\eqref{eq:linearizedMeasurementOperator}.

First we note that, for a general operator $\hat{O}$, we have:
\begin{align}
	d\langle \hat{O} \rangle = \text{Tr}\left\{ \hat{O} d\hat{\rho}_c \right\} \,,
\end{align}
where $\hat{\rho}_c$ is the conditional quantum state. This calculation is further simplified by our Gaussian state assumption, for which expectation values can be factorized into first-order correlates, e.g.
\begin{align}
\label{eq:Gaussian_factorisation}
	\langle \hat{O}_1 \hat{O}_2 \hat{O}_3 \rangle =& \langle \hat{O}_1 \rangle \langle \hat{O}_2 \hat{O}_3 \rangle + \langle \hat{O}_2 \rangle \langle \hat{O}_1 \hat{O}_3 \rangle \\ \notag &+ \langle \hat{O}_3 \rangle \langle \hat{O}_1 \hat{O}_2 \rangle - 2 \langle \hat{O}_1 \rangle \langle \hat{O}_2 \rangle \langle \hat{O}_3 \rangle \,,
\end{align}
using which we find the evolution of the conditional means
\begin{subequations}
\label{eq:eoms_expectations}
\begin{align}
\label{eq:eom_Xj}	d\langle \hat{X}_j \rangle 	&= \omega_j \langle \hat{Y}_j \rangle dt + 2 \sqrt{\alpha \eta} \sum_l \text{Cov}\left( \hat{X}_j, \hat{X}_l \right) d\xi_l(t) \,, \\
	d\langle \hat{Y}_j \rangle 	&= -\omega_j \langle \hat{X}_j \rangle dt + 2 \sqrt{\alpha \eta} \sum_l \text{Cov}\left( \hat{Y}_j, \hat{X}_l \right) d\xi_l(t) \,,
\end{align}
\end{subequations}
and the one-body correlators
\begin{subequations}
	\label{eq:eoms_1bodycorr}
	\begin{align}
		d \langle \hat{X}_j \hat{X}_k \rangle	&= \left( \frac{\omega_j}{2}\langle \hat{X}_k \hat{Y}_j + \hat{Y}_j \hat{X}_k\rangle + \frac{\omega_k}{2}\langle \hat{X}_j \hat{Y}_k + \hat{Y}_k \hat{X}_j\rangle \right) dt \notag \\
								& + 2 \sqrt{\alpha \eta} \sum_l \left( \langle \hat{X}_j \rangle \text{Cov}\left( \hat{X}_k, \hat{X}_l \right) + \langle \hat{X}_k \rangle \text{Cov}\left( \hat{X}_j, \hat{X}_l \right) \right) d\xi_l(t) \,, \label{eq:eom_XjXk}\\
	d \langle \hat{Y}_j \hat{Y}_k \rangle	&= -\left( \frac{\omega_j}{2}\langle \hat{X}_j \hat{Y}_k + \hat{Y}_k \hat{X}_j\rangle + \frac{\omega_k}{2}\langle \hat{X}_k \hat{Y}_j + \hat{Y}_j \hat{X}_k \rangle \right) dt +  \alpha \mathcal{M}_{jk} dt \notag \\
								& + 2 \sqrt{\alpha \eta} \sum_l \left( \langle \hat{Y}_j \rangle \text{Cov}\left( \hat{Y}_k, \hat{X}_l \right) + \langle \hat{Y}_k \rangle \text{Cov}\left( \hat{Y}_j, \hat{X}_l \right) \right) d\xi_l(t) \,, \\
	d \langle \hat{X}_j \hat{Y}_k \rangle	&= \left( \omega_j \langle \hat{Y}_j \hat{Y}_k \rangle - \omega_k \langle \hat{X}_j \hat{X}_k \rangle \right) dt \notag \\
								& + 2 \sqrt{\alpha \eta} \sum_l \left( \langle \hat{X}_j \rangle \text{Cov}\left( \hat{Y}_k, \hat{X}_l \right) + \langle \hat{Y}_k \rangle \text{Cov}\left( \hat{X}_j, \hat{X}_l \right) \right) d\xi_l(t) \,.\label{eq:eom_XjYk}
	\end{align}
\end{subequations}

Furthermore, the Gaussian factorisation Eq.~(\ref{eq:Gaussian_factorisation}) only affects the innovations terms for Eqs.~(\ref{eq:eom_XjXk})-(\ref{eq:eom_XjYk}). Finally, since the RHS of Eq.~(\ref{eq:eom_XjYk}) is real, this implies that 
\begin{equation}
	d \langle \hat{Y}_k \hat{X}_j \rangle = \left( d \langle \hat{X}_j \hat{Y}_k \rangle \right)^\dag = d \langle \hat{X}_j \hat{Y}_k \rangle.
\end{equation}
The equations of motion for the symmetrised covariances are obtained by applying Ito's product rule:
\begin{align}
	d\text{Cov}(\hat{O}_1,\hat{O}_2) = \frac{1}{2}d\langle \hat{O}_1 \hat{O}_2 \rangle + \frac{1}{2}d\langle \hat{O}_2 \hat{O}_1 \rangle - d\langle \hat{O}_1\rangle \langle \hat{O}_2 \rangle - \langle \hat{O}_1\rangle d \langle \hat{O}_2 \rangle - d\langle \hat{O}_1\rangle d\langle \hat{O}_2 \rangle \,. \notag
\end{align}
Equations~(\ref{eq:eoms_1bodycorr}), and the correlation of $d\xi_l(t)$, therefore give
\begin{subequations}
\label{eq:eom_covariances}
\begin{align}
	\frac{d}{dt} \text{Cov}(\hat{X}_j,\hat{X}_k)	&= \omega_j \text{Cov}(\hat{X}_k,\hat{Y}_j) + \omega_k \text{Cov}(\hat{X}_j,\hat{Y}_k)  \nonumber \\   &- 4 \alpha \eta \sum_{l,l'} \text{Cov}(\hat{X}_j,\hat{X}_l) \mathcal{M}_{l l'} \text{Cov}(\hat{X}_{l'},\hat{X}_k)\,, \\
	\frac{d}{dt} \text{Cov}(\hat{Y}_j,\hat{Y}_k)	&= -\omega_j \text{Cov}(\hat{X}_j,\hat{Y}_k) - \omega_k \text{Cov}(\hat{X}_k,\hat{Y}_j) + \alpha \mathcal{M}_{jk}  \nonumber \\  &- 4 \alpha \eta \sum_{l,l'} \text{Cov}(\hat{Y}_j,\hat{X}_l) \mathcal{M}_{l l'} \text{Cov}(\hat{X}_{l'},\hat{Y}_k)\,, \\
	\frac{d}{dt} \text{Cov}(\hat{X}_j,\hat{Y}_k)	&= \omega_j \text{Cov}(\hat{Y}_j,\hat{Y}_k) - \omega_k \text{Cov}(\hat{X}_j,\hat{X}_k)  \nonumber \\  &- 4 \alpha \eta \sum_{l,l'} \text{Cov}(\hat{X}_j,\hat{X}_l) \mathcal{M}_{l l'} \text{Cov}(\hat{X}_{l'},\hat{Y}_k)\,.
\end{align}
\end{subequations}

These results can be recast in matrix form to obtain Eq.~\eqref{eq:GaussianEvolution}.

\section{Approximation scheme for PSDs \label{app:PSD_ApproxScheme}}
In this appendix we develop an approximation scheme for computing integrals of the form
\begin{align}
	\int_{-\infty}^\infty d\omega |\chi_j(\omega)|^2 F(\omega) \,,
\end{align}
where $\chi_j(\omega) = \left[\omega^2-\omega_j^{\rm eff}(\omega)^2 - i\omega \gamma_j(\omega) \right]^{-1}$ is the effective mechanical susceptibility of mode $j$ (\emph{c.f.} Eq.~\eqref{eq:MechanicalSusceptibility}) and $F(\omega)$ is some function that scales as $|\omega|^{\alpha}$ away from resonance for $\alpha > 0$. For $\alpha<4$, the asymptotic behaviour of the integrand is dominated by $|\chi_j(\omega)|^2$, which decays as $\omega^{-4}$ away from resonance. In this case, the integral can be approximated by evaluating $F(\omega)$ at the mechanical resonance, i.e.
\begin{align}
\label{eq:ConvergenceIntegral_Susceptibility}
	\int_{-\infty}^\infty d\omega |\chi_j(\omega)|^2 F(\omega) &\approx F(\omega_j)\int_{-\infty}^\infty d\omega |\chi_j(\omega)|^2 \,.
\end{align}
To analytically compute the integral on the RHS of this expression, we will employ an approximate form of the effective mechanical susceptibility wherein we neglect the shift to the collective mode frequency due to the feedback, and take the damping rate to be its resonant value (\emph{c.f.} Eq.~\eqref{eq:MechanicalSusceptDef}):
	\begin{align}
	\label{eq:MechanicalSuscept_Approximate}
	\chi_j(\omega) &\approx \left[\omega^2-\omega_j^2 - i\omega \gamma_j(\omega_j) \right]^{-1} \,.
\end{align}
We further approximate the mechanical susceptibility by taking $[1+\omega_j^2/\Omega_{j}^2]^{-1}\approx 1$ in the cooling rate \eqref{eq:DampingRate_freqdep}, in which case the integral on the RHS of Eq.~\eqref{eq:ConvergenceIntegral_Susceptibility} can be computed as:
\begin{align}
\int_{-\infty}^\infty d\omega |\chi_j(\omega)|^2 &\approx \int_{-\infty}^\infty d\omega \frac{1}{|\omega^2-\omega_j^2-2i\omega_j^2 c_j \sqrt{\Gamma_j} |^2} \\
&=2\int_{0}^\infty d\omega \frac{1}{\omega^4-2\omega^2\omega_j^2+(1+4c_j\sqrt{\Gamma_j})\omega_j^4}\\
&=\frac{\pi }{2 c_j \sqrt{\Gamma_j } \omega_j^3}-\frac{3 \pi  c_j \sqrt{\Gamma_j }}{4 \omega_j^3} + \mathcal{O}[(c_j\sqrt{\Gamma_j})^3] \label{eq:IntegratedSusceptibilityApprox}\,.
\end{align}
We will retain only the leading order term in this expression, as the prefactor in the RHS of Eq.~\eqref{eq:ConvergenceIntegral_Susceptibility}, $F(\omega_j)$, will contains contributions proportional to small parameters, e.g. $\Gamma_j$, $c_j\sqrt{\Gamma_j}$, and $c_j^2$. We have assumed here the optimal control gain satisfies $c_j=\mathcal{O}(\sqrt{\Gamma_j})$, which is the case for the analytic solution given in Eq.~\eqref{eq:ControlGainOptimal}.

The above approximation scheme enables the analytic integration of the quadrature PSDs, with the exception of the bracketed term proportional to $c_j^2\omega^4$ in Eq.~\eqref{eq:SY_WeakMeas_Approx}. This term describes high-frequency contributions to the steady-state dynamics introduced by the control noise, which have significant non-resonant contributions to quadrature PSDs. We can account for this by separating out resonant and non-resonant contributions to the integral of this term over the frequency domain, with the convergence of the latter dictated by the function $f(\omega)=\omega^4/[1+\omega^2/\Omega_{j}^2]$, i.e.
\begin{align}
c_j^2	\int_{-\infty}^\infty d\omega \frac{|\chi_j(\omega)|^2}{2\pi}f(\omega) &\approx \frac{c_j^2 f(\omega_j)}{2\pi}\int_{-\infty}^\infty d\omega |\chi_j(\omega)|^2 + \frac{c_j^2}{2\pi}\int_{-\infty}^\infty d\omega \frac{1}{1+\omega^2/\Omega_{j}^2} \,, \\
\label{eq:HighFrequencySpectraTermIntegral}	&= \frac{c_j\omega_j}{4 c_j \sqrt{\Gamma_j } }-\frac{3  c_j^3 \omega_j \sqrt{\Gamma_j }}{8} + \frac{c_j^2\Omega_{j}}{2} \,,
\end{align}
where we have taken $\omega^4|\chi_j(\omega)|^2\approx 1$ for the non-resonant term in the first line, and substituted in Eq.~\eqref{eq:IntegratedSusceptibilityApprox} in the second. 

\end{document}